\documentclass[onecolumn]{aa} 
\usepackage{graphicx}
\usepackage{txfonts}
\usepackage[T1]{fontenc}
\begin{document}

   \title{The Universe at extreme magnification.}


   \author{J.M. Diego
          \inst{1}\fnmsep\thanks{jdiego@ifca.unican.es}
          }

   \institute{Instituto de F\'isica de Cantabria (CSIC-UC). Avda. Los Castros s/n. 39005 Santander, Spain
             }


 \abstract{
Extreme magnifications of distant objects by factors of several thousand have recently become a reality. Small very luminous compact objects, such as supernovae (SNe), giant stars at $z=$\,1\,--\,2, Pop III stars at $z>7$ and even gravitational waves from merging binary black holes near caustics of gravitational lenses can be magnified to many thousands or even tens of thousands thanks to their small size. We explore the probability of such extreme magnifications in a cosmological context including also the effect of microlenses near critical curves. We show how a natural limit to the maximum magnification appears due to the presence of microlenses near critical curves. We use a combination of state of the art halo mass functions, high-resolution analytical models for the density profiles and inverse ray tracing to estimate the probability of magnification near caustics. We estimate the rate of highly-magnified events in the case of SNe, GW and very luminous stars including Pop III stars. Our findings reveal that future observations will increase the number of events at extreme magnifications opening the door not only to study individual sources at cosmic distances but also to constrain compact dark matter candidates. 
  }
   \maketitle
%

\section{Introduction}
When observing distant objects with a given instrument, the limiting magnitude set by the experimental configuration defines a natural limit to the maximum distance, $D_{max}$, at which an object with a given luminosity can be observed. If some of these objects are magnified by some factor $\mu$ it is possible to observe a similar object further away up to a distance $D_l(z)=\sqrt{\mu}D_{max}$ (ignoring k-corrections). For sufficiently large values of $\mu$, relatively faint objects can be observed at cosmological distances. It is interesting to answer the question of what is the limit on this relation and how far we can see relatively faint objects that otherwise (i.e without magnification) could not be observed. A similar question has been studied extensively in the context of lensed galaxies and to a lesser extent quasars (QSOs) but these will not be the focus of this work. Instead, we focus on bright and very small objects that can be amplified by extreme magnification factors ($\mu>1000$). The probability of having an event magnified by a large factor $\mu$ is well known to scale as $\mu^{-3}$ \citep[see for instance][]{Lee1990,Rauch1991}. In general, the probability of seeing a strongly lensed event depends on several factors; i) first the volumetric density of objects as a function of redshift, ii) the volume element which depends on the redshift and iii) the probability of intersecting a gravitational lens. The maximum magnification at which an object can be observed depends on the redshifts of the lens and background source, the mass and concentration of the lens and the size of the background sources. Maximum magnification is obtained when a background source is touching a caustic (i.e when a source or radius $R$ is at a distance $R$ from the caustic). The smaller the source, the closer it can get to the caustic so smaller sources can be magnified by larger factors. Stars, for instance, could in principle be magnified by factors of several million when touching a galaxy cluster caustic \citep{Jordi1991}. \cite{Diego2018} discusses how this maximum magnification gets reduced in the presence of microlenses \citep[see also][]{Kayser1986}. The combined effect of macro and microlensing can result in large magnification factors even relatively far away from the critical curve of the macromodel \citep{Diego2018}. However, even with extreme magnifications of several million, a normal star like the Sun at redshift $z>1$ would still fall below the detection limit of the most powerful current telescopes. The background source needs to be not only small, but also very bright. 

In \cite{Kelly2018}, one such very bright object at $z>1$ was found to lie very close to a powerful caustic. 
This star, nicknamed {\it Icarus},  holds the record for the most distant star and most extreme magnification ever observed thanks to gravitational lensing. \cite{Kelly2018} describes Icarus as a $z=1.49$ giant star that is being magnified by the combined effect of a powerful lens (the galaxy cluster MACSJ1149) and a microlens. Both the caustic of the microlens and the background star happen to be aligned near the caustic of the cluster resulting in a very large magnification of a few thousand. This type of observation is the first of its kind and raises the question of how likely this kind of alignments may be. Additional events may have been observed in \cite{Rodney2018} and more recently in \cite{Chen2019}. In \cite{Diego2018}, the authors discuss the interesting possibility of using observations like this one to constrain the amount of dark matter in the form of compact microlenses \citep[see also][where an actual constrain is derived based on Icarus]{Oguri2018}. The mass range that can be constrained this way fills the gap of low-intermediate masses for primordial black holes (or PBH) which have been proposed as one of the candidates to explain dark matter (or at least a fraction of it), as well as the observation of the relatively abundant low frequency LIGO events. In this work we estimate the probability of observing, not only luminous stars at $z>1$ but also discuss other very compact but intrinsically energetic phenomena, such as SNe, or GW.

In order to observe distant gravitationally lensed objects, one needs large magnification factors that compensate for the increase in luminosity distance. The smaller probability of lensing is partially compensated by the larger volume which, between $z\approx 0.3$ and $z\approx 1.3$, grows approximately as $(1+z)^3$ (and at a slower rate at higher redshifts). More precisely, the volume per redshift interval peaks at $z\approx 2.5$ (with about 55 Gpc$^3$ in a redshift interval of thickness $\Delta z=0.1$). Regarding the probability of lensing, background objects at high redshifts have a higher probability of intersecting a gravitational lens along the line of sight. This probability is described by the optical depth. It is well known that the optical depth grows rapidly between z=0 and z=1. Between z=1 and z=3 it continues to grow although at a much slower pace. Beyond $z=$3\,--\,5, the optical depth is still growing but much more slowly and beyond  $z=$5 it only grows by percent values, specially if one considers large magnification factors where the presence of caustics is required (these caustics are expected to be rare for lenses beyond $z\approx 3$). For sources (or events) that trace the star formation history, and at redshifts of the background source between 1 and 3, the small probability of magnification factors can be compensated by the larger volume element and increased volumetric density. In this redshift interval, the probability of seeing extremely magnified events can be maximum. 

Examples of extreme magnification are the aforementioned Icarus event \citep{Kelly2018} with a magnification factor exceeding $2000$. The previous record holder (to the best of our knowledge) was OGLE-2008-BLG-279, for which \cite{Yee2009} infers a magnification of $\approx 1600$. However, this event took place within the boundaries of our Galaxy, {\it not} at cosmological distances. \cite{Zackrisson2015} studies large magnifications in the context of compact globular clusters at high redshift as background objects, and uses N-body simulations to estimate the probability of lensing for a given magnification. They find that a survey with a limiting magnitude of 28 (in the AB system) could find of order 1 primordial globular clusters per 100 deg$^2$ at $z>7$ magnified by a factor $\mu>300$. No observations of globular clusters at cosmological distances lensed by factors of hundreds have been reported yet. 

Earlier work has considered the probability of observing distant very luminous and compact objects, like SNe or QSOs, through gravitational lensing \citep{Martel2008,Oguri2010a} at large (but moderate in the context of this paper) magnifications (i.e $\mu<100$). In \cite{Broadhurst2018}, the authors argue that the low-frequency gravitational wave events observed by LIGO could be interpreted as gravitationally lensed events originating at $z>1$, instead of the implied low-redshift events.  In order for lensing to work in this case, the intrinsic rate of coalescence events at high redshift needs to be over an order of magnitude larger than previously assumed (however this rate is largely unknown), and the magnifications involved need to be of order $10^2$ or $10^3$. 

At redshifts $z>3$, the probability of observing strongly lensed objects declines, unless the intrinsic volumetric density of background objects or their luminosity compensates for the reduction in lensing probability of the larger factor $\mu$.  In \cite{Windhorst2018}, the authors propose to use JWST to observe the first Pop III stars at $z>8$ through caustic crossing events and with magnifications larger than several thousand. Pop III stars can be very luminous (millions of times the luminosity of the Sun) and are expected to be relatively abundant at $z>10$. \cite{Windhorst2018} proposes to monitor massive gravitational lenses at $z\approx$0.4\,--\,0.8 and estimates that, in the most optimistic scenario, one caustic crossing by a Pop III star could be observed every 3 years after targeting very massive galaxy clusters. Observing Pop III stars through caustic crossing would be very exciting, since it would allow to individually study some of the first stars. However, the question of how likely is it to observe this type of event in any surveyed random portion of the sky (nor necessarily including a massive cluster) is still without an answer. One needs to take into account not only the most massive clusters, but all gravitational lenses that are capable of producing caustics at high redshift.  

On cosmological scales, earlier work has looked at the probability of lensing at moderate magnifications ($\mu\approx\,3\,--\,10$), either using analytical models \citep{Turner1984,Fukugita1992,Kochanek1996} or N-body simulations \citep{Hilbert2007,Takahashi2011}. Based on N-body simulations, \cite{Hilbert2007} compute the probability of lensing for magnifications between 3 and 10, and find that for a source at $z=4$ the probability of being magnified by a factor larger than 10 is $\approx 3\times10^{-5}$ \citep[in a subsequent work this probability was revised towards higher values after including the effect of a smooth distribution of baryons but not microlenses, see ][]{Hilbert2008}. \cite{Takahashi2011}, extends the calculation up to $\mu=100$ and finds that for a source at $z=5$ the probability of being magnified by a factor larger than 10 is $\approx 5\times10^{-5}$, in agreement with the result of \cite{Hilbert2007}. At $\mu=100$ and for a source at $z=5$, the probability (per logarithmic interval) drops by an additional two orders of magnitude. These early studies ignore, however, the role of microlenses near critical curves that can be important, as discussed in \cite{Diego2018}, but see also \cite{Venumadhav2017,Oguri2018}. 
   \begin{figure*}
   \centering
   \includegraphics[width=8cm]{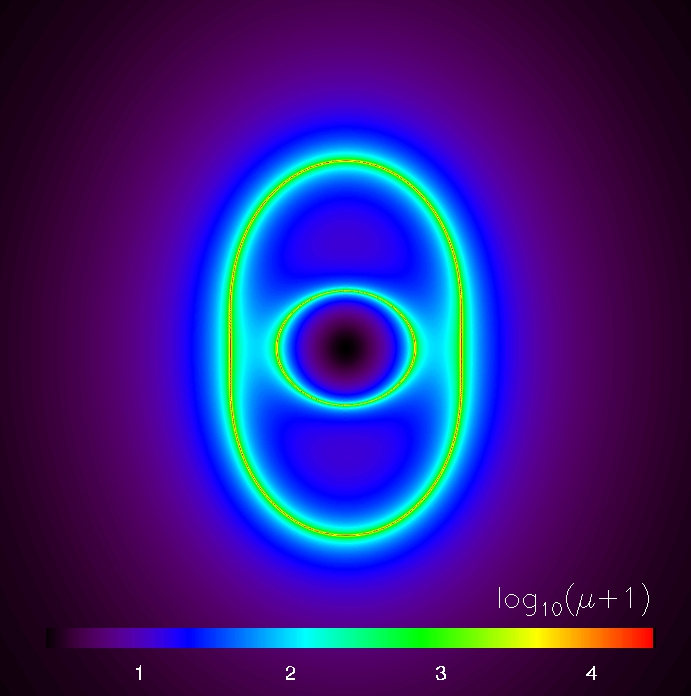}
   \includegraphics[width=8cm]{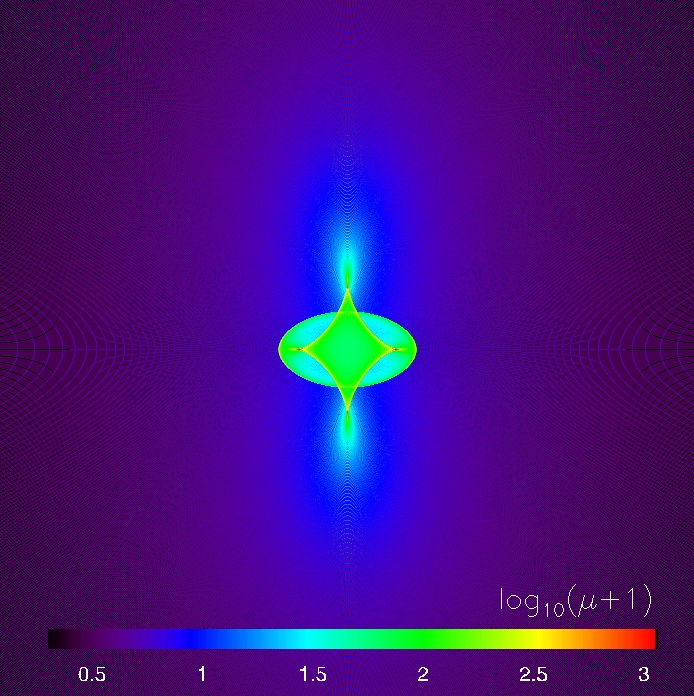}
      \caption{Magnification pattern for a halo at $z_l=0.4$ with mass $2\times10^{15} {\rm M}_{odot}$ and for a source at redshift $z_s=4$. 
               The left panel shows the logarithm of the magnification and critical curves in the image plane (field of view is 2.71 arcminutes across). 
               The right panel shows the logarithm of the magnification and caustics in the source plane (field of view is 1.355 arcminutes across) obtained by 
               inverse ray shooting. 
              }
         \label{Fig_CC_Caustics}
   \end{figure*}

In this work we compute the probability of magnification for bright and compact objects at $z>1$, taking into account the role of microlenses. We rely on an accurate mass function to compute the number density of lenses and an analytical model to map the distribution of mass in individual halos. The analytical model allows us to reach spatial resolutions not available with N-body simulations, and explore the regime of high magnification near the caustics. We focus on very luminous stars (including Pop III stars) and SNe; all of them powerful sources with relatively small radii and which are abundant enough. Their abundance results in non-negligible probabilities for one of them being very close to a caustic. 

We also discuss briefly gravitational waves, although this particular case was explored in \cite{Broadhurst2018}. Here we present in greater detail the lensing probability calculation of \cite{Broadhurst2018} and extend the discussion for the situation where microlenses are present. 

We do not discuss the case of QSO lensing since magnifications in this case are not as extreme as in the case of stars, SNe or GW. The magnifications involved in QSO lensing is normally modest reaching values of $\mu\sim$\,few hundred at most \citep{Walsh1979,Weymann1980,Wang2017}. This limitation in the magnification follows from the much larger size of accretion discs which are orders of magnitude larger than stars or SNe \citep[except for when QSOs are observed at X-ray wavelegths, in which case the size of the X-ray emitting region can be considerably smaller, and comparable to the size of a SNe][]{Mosquera2013}. However, microlensing in QSO plays an important role and the abundant literature on this subject makes fundamental contributions to understand the role of microlenses on the flux fluctuations of multiply imaged QSOs  \citep[see for instance][]{Wambsganss1990}. The smaller size of stars or SNe compared with the size of QSO's accretion discs makes microlensing even more relevant, as the flux fluctuations can be considerably larger than in the case of QSO microlensing. 

The paper is organized as follows; in section \ref{sec_math} we give a brief description of the well known lensing formalism that is relevant for the discussion in this work. In section \ref{section_1lens} we lay the ground with the computation of the lensing probability of a single halo. Section \ref{sect_tau} extends the calculation to a cosmological context where all halos and redshifts are taken into account to compute the lensing optical depth with an emphasis on the optical depth at extreme magnification. Section \ref{sec_micro} discusses the impact of microlenses on the lensing probability. Section \ref{sec_rates} estimates rates of events for super-luminous stars, Pop III stars, SNe and GW. Section \ref{sect_discuss} discusses the results and we conclude in section \ref{sect_concl}. 
We adopt a flat cosmological model with $\Omega_m=0.3$, $\Lambda=0.7$ and $h=0.7$. Using a model with slightly different cosmological values (like the cosmological model inferred by the Planck mission) has no impact on our conclusions since we present order of magnitude estimations of the expected rate of extremely magnified events.  

\section{Formalism}\label{sec_math}

The basic equation in lensing is the so called lens equation that relates the real position in the sky of a background source, $\beta$,  
the apparent position of the observed image, $\theta$, and the deflection angle produced by the lens at that position, $\alpha(\theta,\Sigma)$, 
\begin{equation} \beta = \theta - \alpha(\theta,\Sigma), 
\label{eq_lens} 
\end{equation} 
where $\Sigma(\theta)$ is the surface mass density of the lens at the position $\theta$. 
The (vector) deflection angle is obtained from the derivatives of the (scalar) lensing potential, 
\begin{equation}
\psi(\theta) = \frac{4 G D_{l}D_{ls}}{c^2 D_{s}} \int d^2\theta'
\Sigma(\theta')ln(|\theta - \theta'|), 
\label{2-dim_potential} 
\end{equation}
where $D_l$, $D_s$, and $D_{ls}$ are the angular diameter distances to the lens, to the source and from the lens to 
the source, respectively. Given the surface mass density, $\Sigma$, one can compute the lensing potential and its derivatives to obtain the deflection field. 
Once the deflection field is known, the magnification in the image plane can be derived as a combination of derivatives of the deflection field, and through the lens equation. The magnification in the source plane can be computed using the inverse ray shooting technique. In the image plane, regions where the magnification diverge are known as critical curves. The corresponding mapping of these curves into the source plane (through the lens equation) produce a different set of curves known as caustics. Near a critical curve, one can Taylor expand the lens equation and after retaining only the first orders find that 
\begin{equation}
|\beta-\beta_o| = \frac{|\theta-\theta_0|^2}{\Theta}
\end{equation}
where $\Theta$ is a constant (usually expressed in arcseconds) that depends on the lens model and redshifts of the lens and source, $\beta_o$ is the position of the caustic and $\theta_o$ is the corresponding position of the critical curve. As a consequence, near a critical curve, the magnification falls as the inverse of the distance to the critical curve,  
\begin{equation}
\mu = \frac{\mu_o}{|\theta-\theta_o|}, 
\end{equation}
where $\mu_o$ (also expressed usually in arcseconds) is a constant that depends on the slope of the lens potential at the position of the critical curve (shallower potentials result in larger values of $\mu_o$). Later in the paper we will refer to the tangential magnification, $\mu_t$, and radial magnification, $\mu_r$, with the total magnification being the product of the two, $\mu=\mu_t\mu_r$.
In the source plane, the magnification then falls like 
\begin{equation}
\mu = \frac{\mu_o/\sqrt{\Theta}}{\sqrt{|\beta-|\beta_o|}}
\label{Eq_MuBeta}
\end{equation}
Finally, the probability of having magnification larger than a certain value $\mu$ is equal to the probability of being at a distance to the caustic less than the corresponding $\Delta\beta=|\beta-\beta_o|$, i.e, $P(>\mu) \propto \Delta\beta \propto \mu^{-2}$. The differential probability is then $dP(\mu)/d\mu\propto\mu^{-3}$. 
From expression \ref{Eq_MuBeta}, the normalization factor $\mu_o/\sqrt{\Theta}$ can range between $\sim$1 for small halos to 20\,--\,40 for the most massive clusters. For intermediate lenses with normalization $\mu_o/\sqrt{\Theta} \approx 10$, a source at a distance of 1 milliarcsec (typically a few parsecs at $z>1$) would have a total magnification of $\approx 300$. 

\section{A single lens}\label{section_1lens}
The above discussion is for regions that are close to a caustic. In this subsection we explore the magnification properties of a single lens but for all possible distances and magnifications. Even though most of the results in this subsection are widely known, it will be instructive for the remaining of the paper, in particular to understand the shape of the probability of lensing at intermediate magnifications and the difference between computing the magnification in the source plane with the inverse ray shooting method or the deprojection method. 
 
We assume a cored elliptical Navarro, Frenk and White (or NFW) profile \citep{Navarro1997} with ellipticity $e=0.2$, where an NFW profile symmetric in $r=\sqrt{x^2+ y^2}$ is modified after substituting $r$ by $r' = \sqrt{x^2/(1-e) + y^2(1+e)}$. In the central part of the NFW halo, we assume the profile goes like $(r+ r_c)^{-1}$ instead of the usual $r^{-1}$ dependency where  the core radius scales with mass as $r_c=20({\rm M}_{vir}/10^{15})$ kpc. The core radius avoids unphysical divergences at $r=0$ but also reproduces the profile of many massive lenses where the BCG usually exhibits a core, or flattened Sersic profile (i.e with Sersic index $n\approx 1$). Adopting a standard NFW profile with no core should have minimal impact in our calculations, specially at large magnifications, where most of the contribution to the lensing probability comes from regions near the tangential critical curves, that is, far from the core region. We should note that for small halos, the presence of the core can reduce their contribution to the optical depth since the central region of some of these halos may be subcritical when a core is present. The contribution of the small halos to the optical depth is discussed in more detail in the following section. 
For the virial radius, we adopt the scaling  $R_{vir}=1.5*(M_{vir}/10^{15})^{1/3}$ Mpc, and concentration, $C$, given by the model in \cite{Prada2012}. 
\begin{equation}
 C = 7.28\left(\frac{M_{vir}}{10^{12}}\right)^{-0.074}
\end{equation}
We note that in the previous equation, we ignore the evolution with redshift. Since most of the lenses concentrate around redshift 0.5, we expect this to be a small effect, specially for massive halos for which the dependency of the concentration with redshift is weaker \citep[see for instance Figure 12 in][]{Prada2012}.
The ellipticity $e=0.2$ is typical of halos. Different ellipticities have a small impact on the derived probability of lensing (provided the concentration remains constant). The concentration, on the other hand, plays a more significant role. Halos that are more concentrated increase the probability of lensing, specially of large magnifications. Our results should be considered conservative as we are ignoring projection effects  (and the effect of baryons) that could increase the concentration of a halo (and consequently the probability of lensing) if it is aligned in the line of sight, or two small halos fall in the same line of sight.
We simulate halos within a region extending up to 2.3 times the virial radius (a sufficiently large area is needed in order to minimize edge effects). This is sufficient to contain the entire halo when the ellipticity is $e=0.2$. The simulated halos have typical spatial resolutions of $0.6$ kpc per pixel for a lens at $z\approx 0.6$ and mass of $2\times10^{15}{\rm M}_{\odot}$ and about 50 pc for a halo of $10^{12}{\rm M}_{\odot}$ at the same redshift. This resolution is significantly higher than what can be obtained with cosmological N-body simulations. This is important to reproduce with accuracy large magnifications. For the halo profiles, we consider an NFW model. Earlier work focusing on smaller galaxies have used different profiles, with the isothermal model being one of the most popular. However, to explore in detail the regime of extreme magnifications, intermediate mass  halos are the most relevant and for these type of halos, the NFW profile gives a good description of the total mass (dark matter plus baryons) specially for large mass halos.  We note that we explicitly ignore the role of baryons, or the central galaxy in the halo (other than by assuming a core radius in the NFW profile, as described above). Baryons are expected to increase the convergence in the central part of halos (although they can also decrease it due to feedback). This has a non-negligible impact on small halos that may be sub-critical without the help of the baryonic component (large halos are already supercritical (that is, their surface mass density is above the critical surface mass density in parts of the halo) and adding the baryons has a small effect on the probability of magnification. The effect of ignoring the baryon effect is discussed in more detail in the following section. 
   \begin{figure}
   \centering
   \includegraphics[width=9cm]{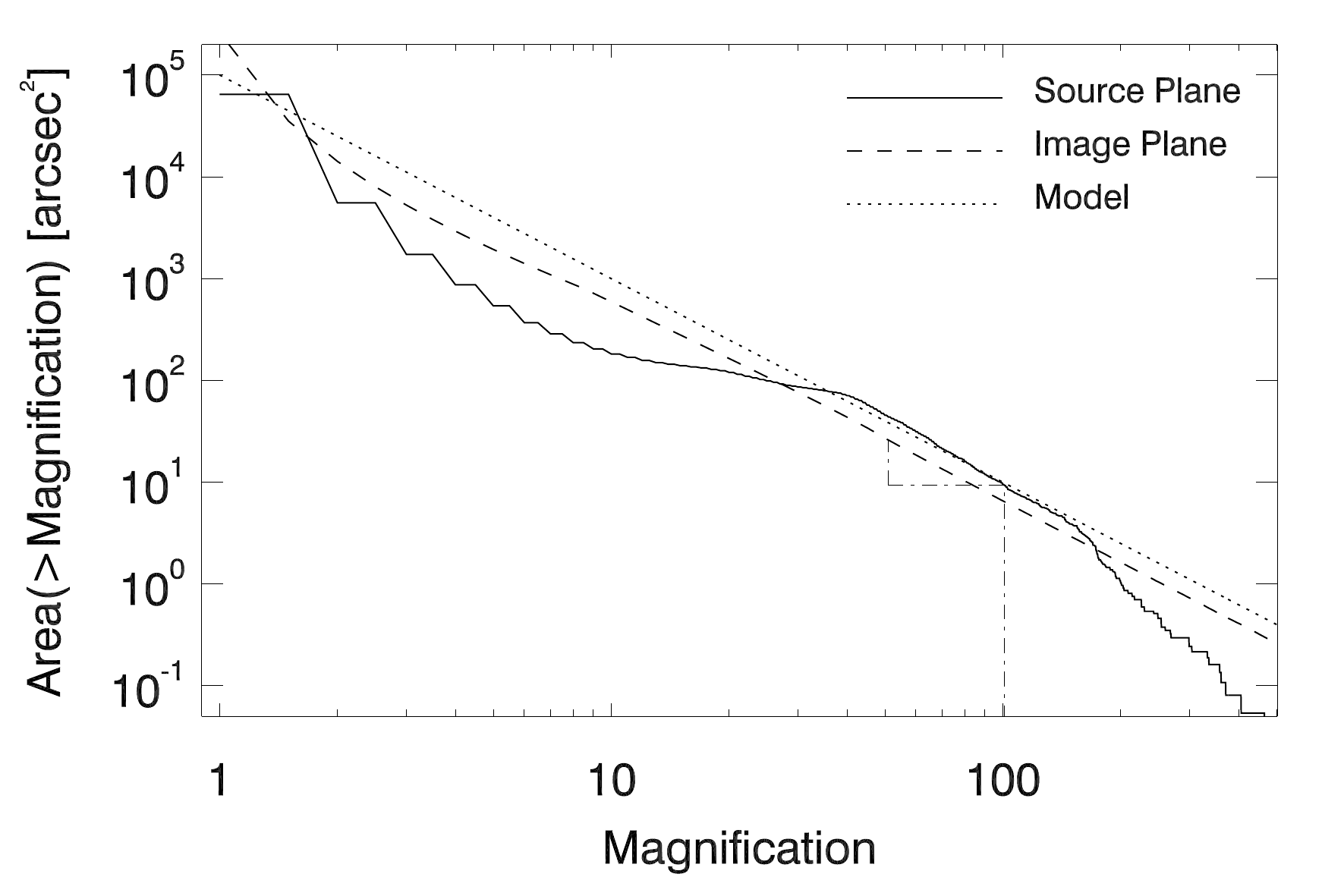}
      \caption{Area above a given magnification for a single halo at $z_l=0.4$ with mass $2\times10^{15} {\rm M}_{\odot}$ and for a source at redshift $z_s=4$. Note how the area computed in the image plane does not follow the $\mu^{-2}$ law at low and moderate magnifications ($\mu<40$). At very high magnifications ($\mu>150$), resolution effects break the $\mu^{-2}$ law but the area can be safely extrapolated to large values of the magnification above $\mu\approx 50$.  
              }
         \label{Fig_AGTmu1}
   \end{figure}

The magnification is computed based on the inverse ray shooting method \citep[see for instance][]{Kayser1986}. 
This method is appropriate to compute the total magnification including all possible counterimages. The total flux needs to be accounted for when considering extreme magnifications, since counterimages may appear separated by a small fraction of an arcsec, and hence be unresolved or barely resolved (or a macroimage may break into multiple microimages separated by a few milliarcseconds or less). 
In Figure~\ref{Fig_CC_Caustics} we show the magnification for a halo at $z_l=0.4$ with mass $2\times10^{15} {\rm M}_{\odot}$ and for a source at redshift $z_s=4$. The left panel shows the central portion of the halo with the magnification computed in the image plane (assuming a point source in the background). The critical curves where the magnification diverges are clearly visible. The color scale indicates the magnification in logarithmic units.  The right panel shows the magnification in the source plane as derived from the inverse ray shooting technique. The field of view is in this case half the size the field of view in the left panel. The caustics are also clearly visible with the diamond shape caustic mapping into the tangential critical curve and the elliptical caustic mapping into the radial critical curve. In the context of this paper, the most striking aspect that is of interest to this work is the clear discontinuity in the magnification at the caustics. Contrary to what happens with the magnification in the image plane, where the magnification appears to be continuous across the image, the source plane shows a sharp discontinuity at the caustics. This fact has interesting implications for the probability of large magnification. To compute the probability of magnification, two different approaches are commonly used in the literature. The fast {\it deprojection method} is used normally when computing lensing probabilities for large lenses, like galaxy clusters. This is an inexpensive method that simply deprojects the image plane (using the lens equation) into the source plane. Since the magnification, $\mu$, is defined as the ratio of areas between the observed images and the sources, the area of a pixel with magnification $\mu$ in the image plane is reassigned to an area in the source plane that is $\mu$ times smaller. This method is very fast but does not give the right probability for the total magnification by ignoring the multiplicity of images.  When multiple images are produced, several regions in the image plane (with different magnifications) project back into the same area in the source plane.

Instead, an {\it inverse ray shooting method}, where pixels that originate in the image plane and land in the source plane are counted for every pixel in the source plane, is the appropriate method to compute the total magnification. In this work we are interested only in the total magnification, since we assume that counterimages of a small background source with extreme magnification form very close to the critical curve and the multiple images (carrying the total magnification) may appear as a single unresolved image.  Figure~\ref{Fig_AGTmu1} shows the difference between the two methods for the cluster shown in Figure~\ref{Fig_CC_Caustics}. The dashed line corresponds to the case where the area is computed in the image plane and later divided by $\mu$ (deprojection method). This method overestimates the area at low and intermediate magnifications and underestimates it at high magnifications. The inverse ray tracing method is shown as a solid line  in Figure~\ref{Fig_AGTmu1}. The inverse ray shooting method suffers of resolution limitations at very large magnifications. In this regime, the width of the region around the caustic that has such large magnification is smaller than the pixel size used to do the mapping between the image and source planes. As a consequence, the probability of lensing does not follow the $\mu^{-2}$ law at the highest magnification. This can be seen in the solid line at magnifications larger than $\mu\approx150$. 

%
   \begin{figure*}
   \centering
   \includegraphics[width=9cm]{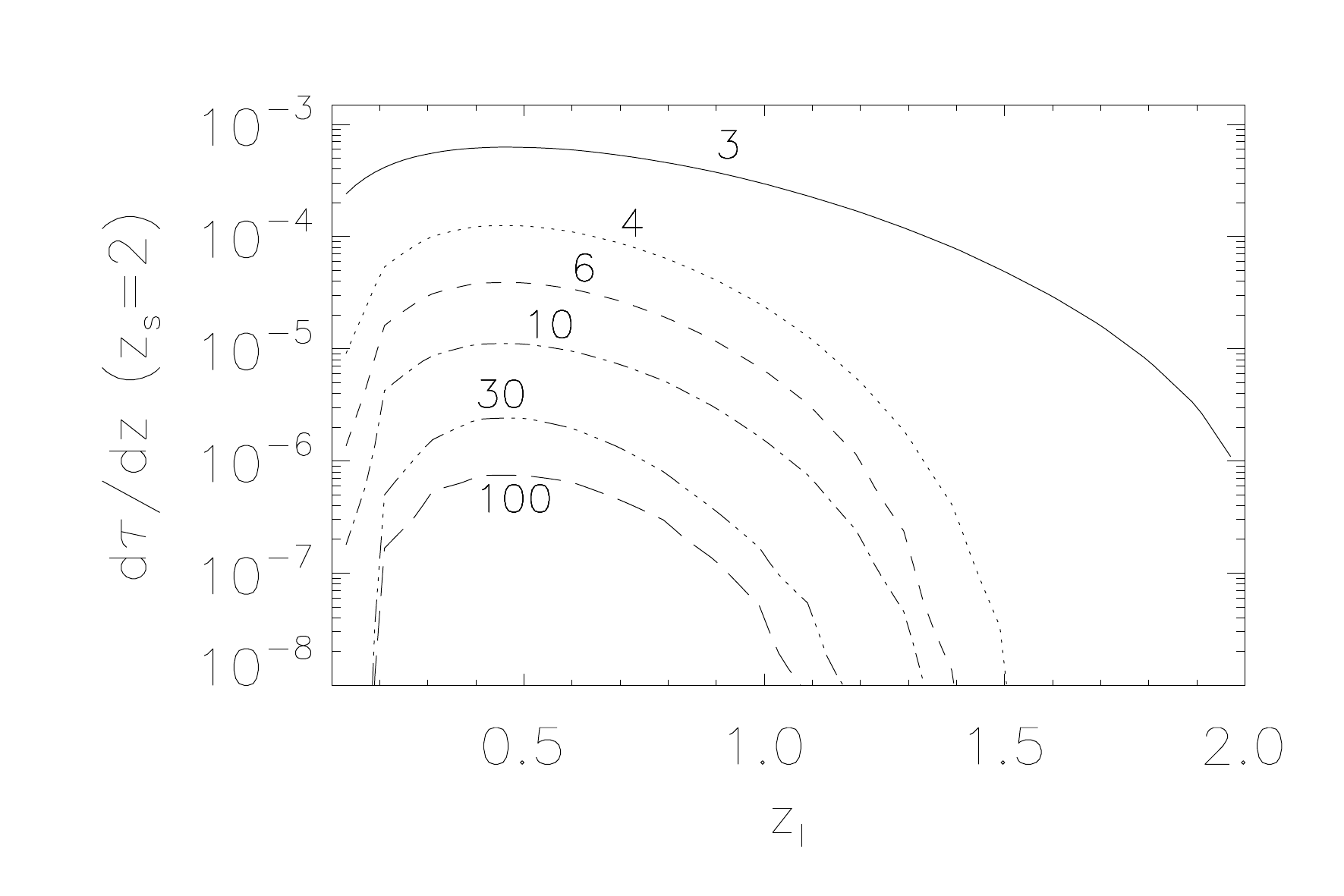}
   \includegraphics[width=9cm]{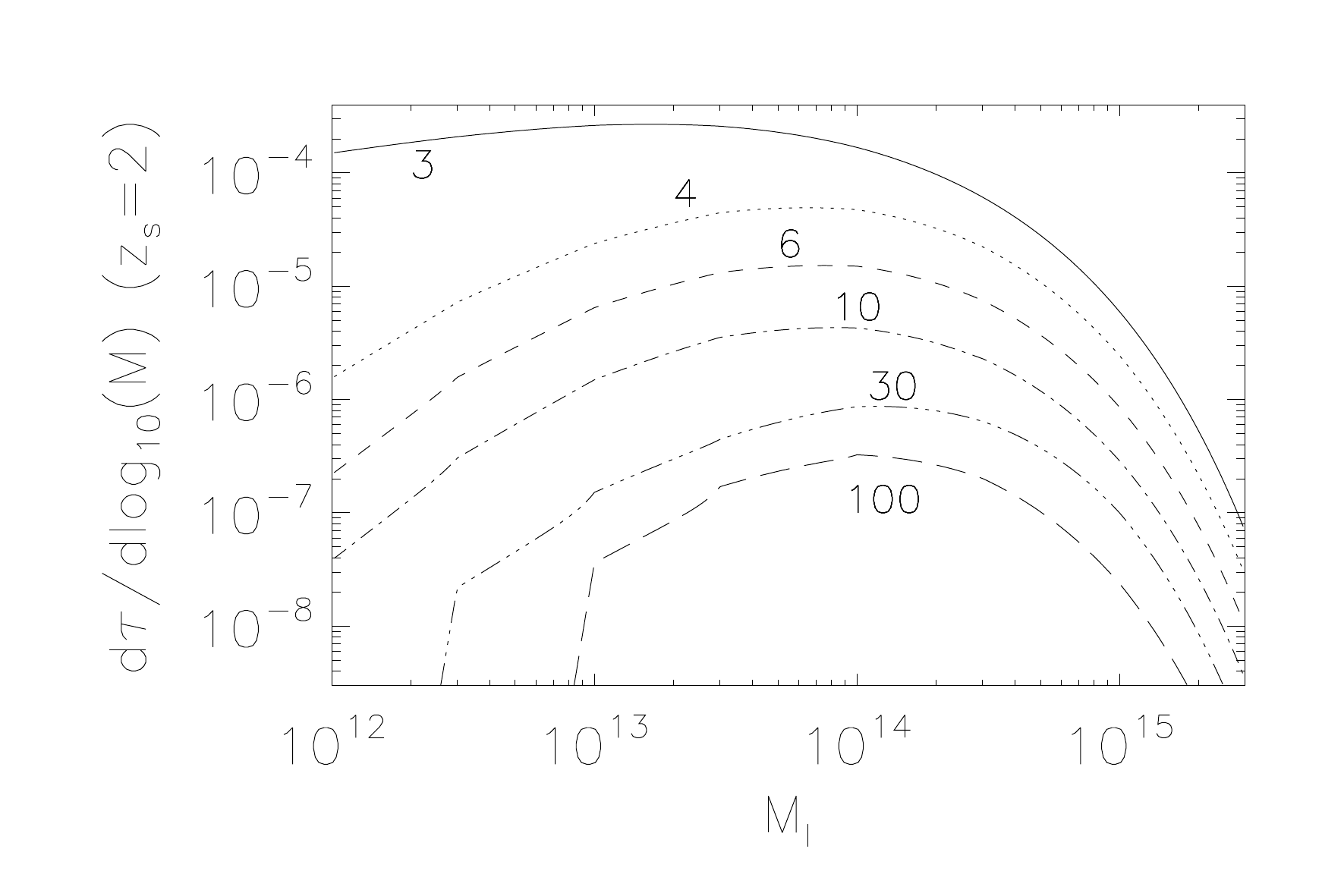}
      \caption{Left panel. Contribution to the optical depth for a source at redshift $z_s=2$ as a function of the redshift of the lens $z_l$ and for all lens masses.  
               Right panel. Contribution to the optical depth for a source at redshift $z_s=2$ as a function of the mass of the lens $M_l$ and for all lens redshifts. In both cases, the numbers by each curve indicate the magnification that the curve contributes to. Note how for large magnifications only the most massive halos contribute to the magnification.   
              }
         \label{Fig_Zl_Ml}
   \end{figure*}

The remedy for this problem is to simply extrapolate the  $\mu^{-2}$ law from this point. The plateau observed at intermediate magnifications ($10<\mu<40$) corresponds to the gap in magnification in the source plane between the regions inside the caustics and the regions outside the caustics. Generally, the larger the caustic region, the larger this gap and the more pronounced the plateau. Above a certain value of the magnification ($\mu\approx40$), the area follows the standard law $\mu^{-2}$ expected for fold caustics. This value of the magnification ($\mu\approx40$) corresponds approximately with the local minimum of the magnification at the center of the diamond shaped caustic. Halos that are subcritical (that is, their surface mass density is smaller than the critical surface mass density) do not have a plateau and fall faster than the $\mu^{-2}$ law without reaching large magnifications. As a halo approaches the critical point, its probability of lensing starts approaching the $\mu^{-2}$ law. 
To better show the relationship between the two methods, in Figure~\ref{Fig_AGTmu1} we show with a dot-dashed line the connection between the two in the regime of high magnification, where both curves fall like $\mu^{-2}$. If one takes a point with magnification $\mu=50$ on the dashed line and divides the area by two moving downwards in the vertical direction and later moves in the horizontal direction up to the point where the magnification is  $2\mu=100$, one ends in the solid curve. In other words, the deprojection method is counting the area twice but it only accounts for the magnification of one of the counterimages. This is easy to understand, since at high magnification, most of the flux is divided into two counterimages, each one carrying approximately half the total magnification. At lower magnifications, there may be 3 or more dominant counterimages making the relationship between the two methods less clear.

\section{Lensing probability for a cosmological volume}
\label{sect_tau}

The probability of having an event with magnification larger than certain value is a key ingredient to compute the rates of lensed events. The number of observed events from the redshift interval $z_s \pm \Delta z/2$ that are lensed with magnification larger than $\mu$  by a population of lenses between redshift 0 and redshift $z_s$ is given by
\begin{equation}
\frac{dN(>\mu,z_s)}{dz_s} = dV(z_s)\,\Delta z\,R(z_s)\,P(>\mu,z_s)
\end{equation}
where $dV(z_s)\Delta z$ is the volume element at $z_s$ contained in the spherical shell with thickness $\Delta z$, 
$R$ is the intrinsic rate of events or number of events per unit volume and unit time at the redshift $z_s$,  and $P(>\mu,z_s)$ 
is the probability of having a magnified event at $z_s$ that is being amplified with magnification larger than $\mu$. When $R$ represents a rate of events (transients), it needs to be divided by the factor $(1+z_s)$ in order to account for the time stretch between the observed rate and the intrinsic rate.  

The product $V(z_l)\,\Delta z\,R(z_s)$ corresponds to the total number of events per unit time in the volume being lensed ($z_s-dz_s < z < z_s+dz_s$). 
The probability $P(\mu,z_s)$ is defined as as 
\begin{equation}
P(>\mu,z_s)=\int_0^{z_s}dz_l\frac{dp(>\mu,z_l,z_s)}{dz_l}
\label{Eq_TauP}
\end{equation}
where $dp(>\mu,z_l,z_s)/dz_l$ is the fraction of the total area (at the distance of the source) that is being magnified by a factor larger than $\mu$ \citep[see for instance][]{Turner1984}
\begin{equation}
\frac{dp(>\mu,z_l,z_s)}{dz} = A_T(z_s)^{-1}\frac{dV(z_l)}{dz_l}\int dM \frac{dN}{dMdz} A_N(\mu,M,z_l,z_s),
\end{equation}
where $dV(z_l)/dz_l$ is the volume element at the redshift of the lens integrated over the entire 
sky, $dN/dMdZ$ is the halo mass function, 
$A_N(>\mu,M,z_l,z_s)$ is the area with magnification larger than $\mu$ for a lens with mass $M$ at redshift $z_l$ lensing a 
background source at redshift $z_s$ and $A_T(z_s)$ is the area in physical units of the spherical shell at redshift $z_s$. That is,  
\begin{equation} 
A_T(z_s) = 4\pi D_a(z_s)^2, 
\end{equation}
where $D_a(z_s)$ is the angular diameter distance at the redshift of the source. 
The definition of $P(>\mu,z_s)$ is essentially the same as the optical depth of lensing by a factor larger than $\mu$, so we can refer to $P(>\mu,z_s)$ as $\tau(>\mu,z_s)$. 

For the mass function, we consider the one from \cite{Watson2013} which is appropriate for a wide mass range (from small groups to the most massive clusters). 
Alternative mass functions like the Tinker mass function are expected 
to produce similar results, since the main differences between these mass functions appear at the regime of very massive cluster which have a small impact for our study. The \cite{Watson2013} mass function is more accurate than Tinker's in this regime, due to the larger volume of the JUBILEE simulation \citep{Watson2014}. We impose a lower cutoff in the mass and compute the integral between the mass limits $M_{min}=10^{12} M_{\odot}$ and $M_{max}=3\times10^{15} M_{\odot}$. 
The selected mass range is appropriate for computing the probability of lensing at high magnification as we show in section \ref{sec_tau2}. At low magnifications, the mass range needs to be extended toward lower masses, since even subcritical halos can contribute in this regime. However, a lower mass limit of $10^{12} M_{\odot}$ is sufficient for our purposes since we focus on the high magnification regime. To model the quantity $A_N(>\mu,M,z_l,z_s)$ we simulate caustics for different halo masses, halo redshifts and background sources, then compute $A_N(>\mu,M,z_l,z_s)$ for each model as described in the previous section through inverse ray tracing. Since inverse ray tracing has limited resolution in ares of high magnification, we compute the lensing probability $P(>\mu,z_l,z_s)$ only up to values of $\mu=100$. At values of $\mu$ significantly larger than $\mu=100$, we observe resolution effects. However, $\mu=100$ is sufficiently large that above this number,  $P(>\mu,z_l,z_s)$ scales as the usual $\mu^{-2}$ law, typical of fold caustics, and can be safely extrapolated for each halo towards higher magnifications. Obviously, this extrapolation is performed only for those halos that exhibit supercritical behavior (i.e have caustics). For subcritical halos no extrapolation is needed as they simply do not reach the value $\mu=100$. This computation is repeated for different halo masses, halo redshift and source redshift. We adopt an adaptive resolution scheme, so the magnification around small halos are computed at higher resolution than larger halos in order to properly resolve the caustic regions. The global probability of lensing of a source at redshift $z_s$  is computed after integrating the lensing probabilities of all individual halos in the mass range considered, and up to the redshift of the source. In this work we assume a maximum redshift for the lenses of $z_{max}=2.0$, since most lenses that contribute to the high magnification are below $z=2$. For the sources, we assume $z_{max}=5.0$.  
The increase in probability between redshift $z_s=5$ and $z_s=10$ is modest, at the level of 20\% \citep{Zackrisson2015}.

As mentioned earlier, the optical depth, $\tau(>\mu,z_s)$, 
is given by Eq.~\ref{Eq_TauP}. 
The optical depth defined this way gives the probability of magnification with a factor larger than $\mu$ of a given background source at $z_s$ with infinitesimally small radius.

\subsection{Optical depth at high magnification}\label{sec_tau2}
   \begin{figure}
   \centering
   \includegraphics[width=9cm]{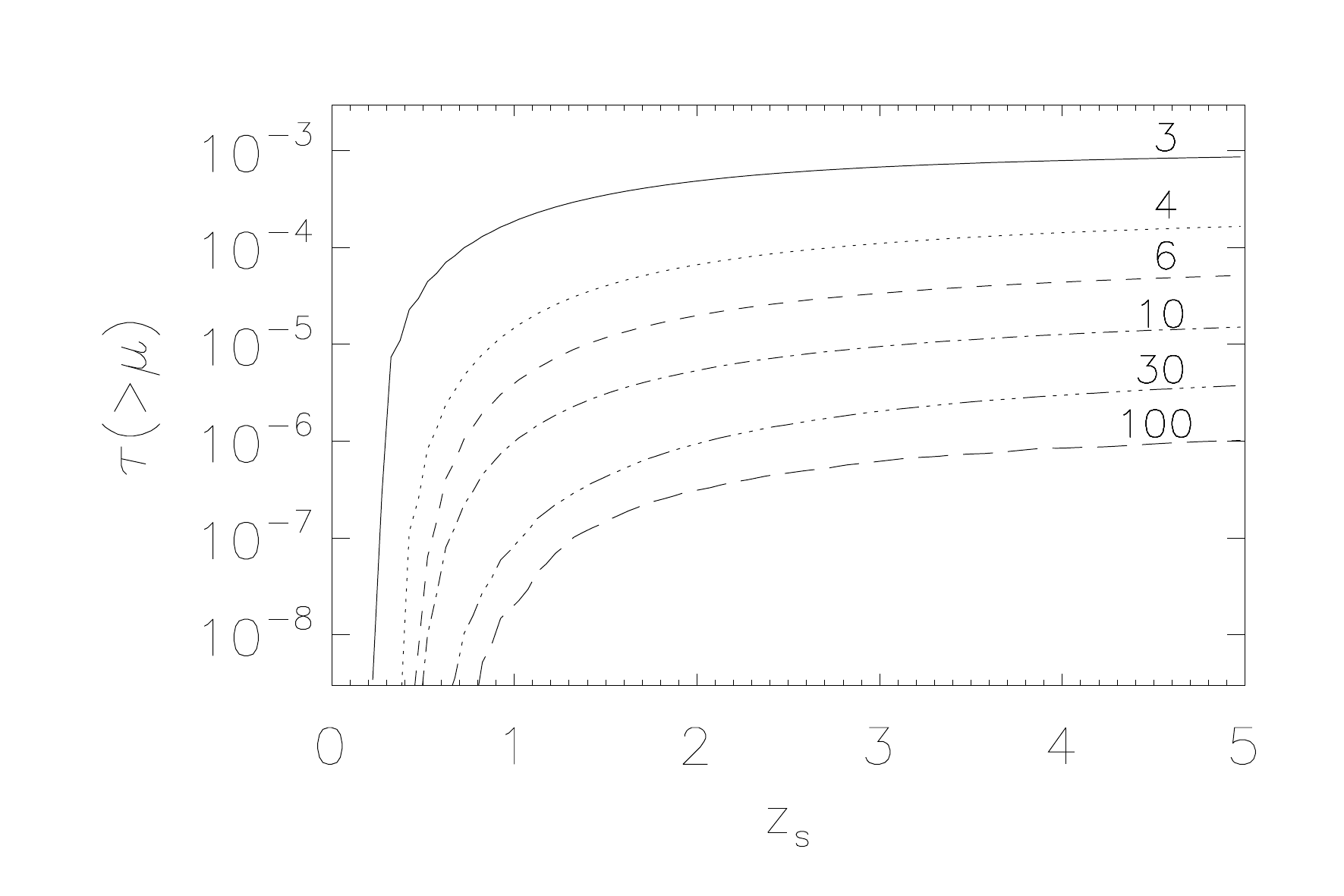}
      \caption{Optical depth for different magnifications as a function of redshift. The numbers by each curve indicate the corresponding magnification.
              }
         \label{Fig_Tau_Mu}
   \end{figure}

Figure~\ref{Fig_Zl_Ml} shows the differential contribution to the optical depth for a source at  $z_s=2$  as a function of the redshift of the lens (left panel) and as a function of the mass of the lens (right panel).  The numbers next to each curve show the magnification at which the contribution is computed. In the left panel, we have integrated over the entire mass range. At low magnifications, contributions from halos come from the entire redshift range with a peak around $z \approx 0.5$. However, as the magnification increases, the redshift range of the halos that contribute to that magnification gets reduced, although maintaining the peak at $z \approx 0.5$. The reason for this decrease is made more evident in the right panel where now we have integrated over the total redshift range. Large magnifications are produced only by large lenses. The sharp decline in optical depth for $\mu=100$ at $10^{13}$ solar masses marks the transition between subcritical and supercritical halos. At lower magnifications ($\mu<10$) small halos play an important role and can not be neglected. 

The right panel of Figure~\ref{Fig_Zl_Ml} predicts that the halos that contribute the most to the optical depth at high magnification are in the range of $10^{14}$ M$_{\odot}$ (at magnification factors $approx 3$ halos ten times smaller are the main contributors). 
The calculation captures the fact that larger lenses, although less numerous, can magnify more than one background object, as evidenced by galaxy clusters that can lens tens of sources each, while smaller halos typically lens $\sim 1$ background object at most.

This appears to be in tension with current observations, and also some earlier theoretical work that suggest that less massive halos are the most typical lenses \citep[see however Figure 5 in][where a similar result is also  found]{Hilbert2007}. This is in part supported by results from  optical surveys, such as SDSS, where it is found that most known gravitational lenses are less massive halos, \citep{Inada2012}, not the more massive halos expected from Figure~\ref{Fig_Zl_Ml}. 
Earlier work has predicted also the lensing effect from these early type galaxies. There are different reasons responsible for this tension and that deserve a discussion in this section. From the theoretical side, most of the previous calculations found in the literature focus on low magnifications ($\mu\sim3$). In the regime of low magnifications, the role of low mass halos is more relevant as shown in the right panel of Figure~\ref{Fig_Zl_Ml}. At larger magnifications, more massive halos take a leading role as shown by the same figure \citep[see also][where a similar trend is observed also for single highly magnified images]{Keeton2005}. In earlier predictions, it is also customary that the optical depth is computed in the context of galaxy-galaxy lensing which explicitly ignores the contribution to the optical depth from groups and clusters. These calculations draw the population of lenses from the velocity dispersion of early type galaxies \citep{Fukugita1992,Collett2015}. The velocity dispersion of galaxies derived from observations sets a natural limit on the maximum velocity dispersion, at the maximum radius of the stellar component. 
Also, much of the literature explicitly focuses on SIS profiles that are known to increase the optical depth of low mass halos compared with the NFW halos used in our work. See for instance Figures 9-11 in \cite{Lapi2012}, Figure 2 in \cite{Porciani2000}, or Figure 9 in \cite{Perrotta2002}  where the dependency with M is explicitly shown. 
Regarding observations, the discrepancy between our calculation and observations (where most strong lenses are expected to be smaller halos) is mostly due to the fact that we ignore the (cooling) role played by baryons  since they can increase the relative contribution of small halos to the optical depth of lensing. Baryons can promote the subcritical central region of a small halo to supercritical values. This is in agreement with observations, where most QSOs and lensed galaxies have been found so far around early type galaxies in optical surveys, such as SDSS \citep{Inada2012}, and not in more massive halos as predicted by our calculation. However, the observed magnifications for these lensed QSO are moderate (a few tens at most). For extreme magnifications, larger halos are more relevant and for these massive halos, the role of the baryonic component, or the central galaxy, diminishes since these halos are already supercritical \citep{Meneghetti2003}.

Large halos are better modeled by NFW profiles, while smaller halos are better represented by SIS profiles. \cite{Kochanek2001} argues that around the galaxies embedded in halos, the change from NFW to SIS "explains why many lenses found in groups of
galaxies  were  associated  with  the  galaxies  in  the  group rather than the group halo, even though the group halo had to be more massive than its component galaxies".
It is unclear to what extent many of the elliptical galaxies identified as lenses in optical surveys, such as SDSS, are not in reality the central galaxy of more massive halos with virial masses several times $10^{13}$ M$_{\odot}$ (some are clearly identifiable as such). Early type galaxies with large velocity dispersion are known to be at the centres of halos with several times $10^{13}$ M$_{\odot}$. 

A more accurate modelling would take into account this transition between the low mass halos (SIS-like) to the more massive halos (NFW-like) \citep[see for instance][]{Keeton1998,Porciani2000}. This transition is mostly due to the more important role that baryons (cooling) play in small halos \citep{Kochanek2001}. \cite{Porciani2000} shows how below a halo mass scale of $\approx 3.5\times 10^{13}$ an SIS model is more appropriate while for masses above this value, an NFW profile is a better description. A hybrid model where small substructures (including the central galaxy) are modeled as SIS while the main halo follows an NFW-like profile may be the most suitable models. Such detailed modelling is, however, beyond the scope of this paper. Since this work focuses on the largest magnifications, for which more massive halos are more relevant, we consider for simplicity only NFW profiles but the reader should be aware that the contribution from small halos or substructures (specially at low magnifications) is understimated by adoptiong an NFW profile. Our results on the optical depth  should then be considered conservative.

\begin{figure}
   \centering
   \includegraphics[width=9cm]{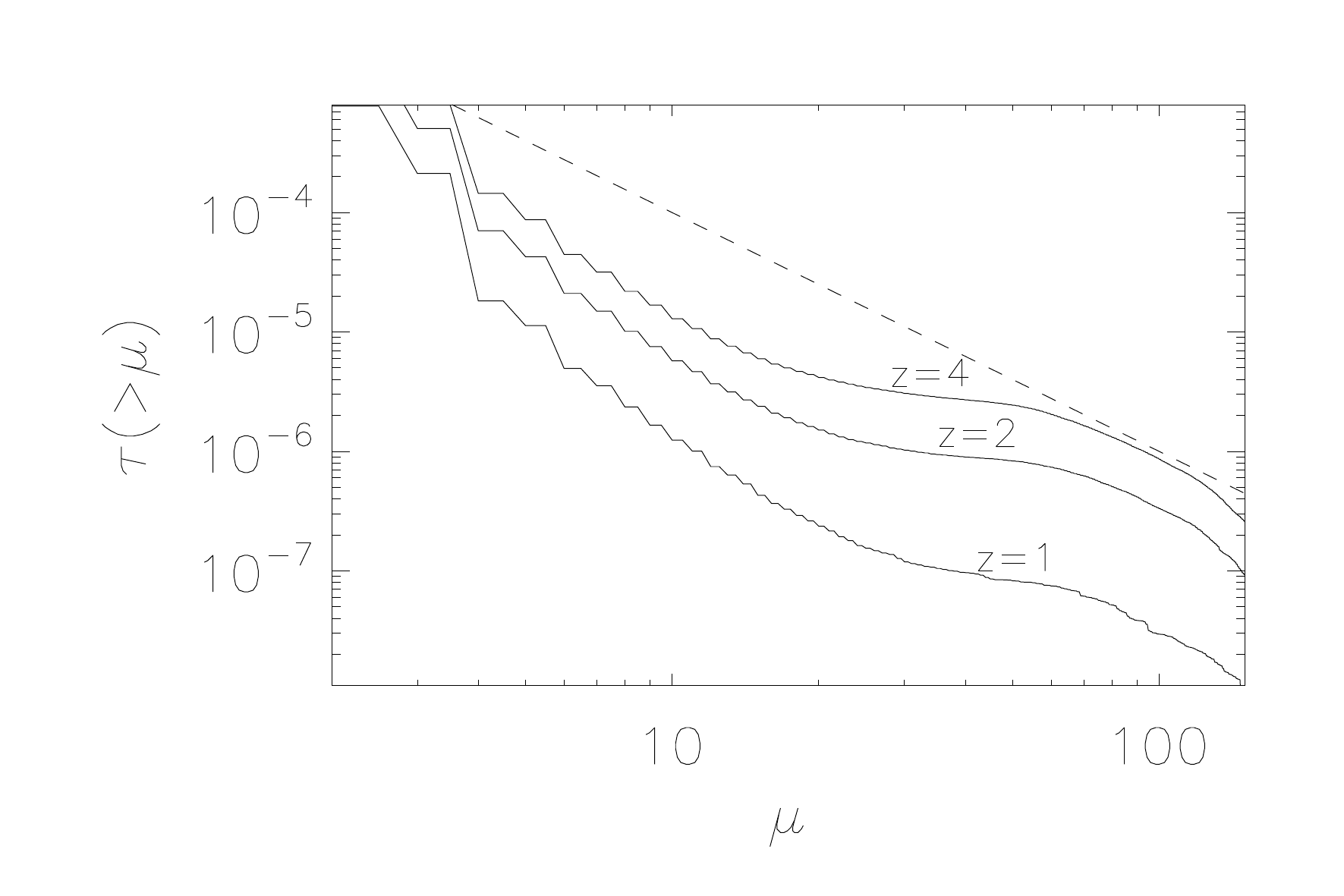}
      \caption{Optical depth for different redshifts as a function of magnification. The dashed line is the power law $\mu^{-2}$. 
              }
         \label{Fig_Tau_Mu_Z}
   \end{figure}

Figure~\ref{Fig_Zl_Ml} also shows clearly that the integrated optical depth decreases rapidly with increasing magnification. This is shown in more detail in Figure~\ref{Fig_Tau_Mu} where again the number next to the curves indicates the magnification. As found in earlier work, the probability of lensing declines sharply below $z_s=1$ and changes much more gently above $z_s=2$. The relative change between $\mu=3$ and $\mu=10$ is much more abrupt than between  $\mu=10$ and $\mu=30$ or  $\mu=30$ and $\mu=100$. This is a direct consequence of the transition between the subcritical and supercritical regime shown in Fig.~\ref{Fig_AGTmu1}, where the probability of lensing drops faster than $\mu^{-2}$ at low magnification factors. The same pattern is observed when integrating over cosmological volumes as shown in Figure~\ref{Fig_Tau_Mu_Z}. Note how in this plot, the $\mu^{-2}$ is satisfied at $\mu<100$ for all redshifts. 

The optical depth is, in general, in agreement with previous results based on N-body simulations \citep[see Figure 5 in][]{Hilbert2007}. Although it should be noted that the optical depth in \cite{Hilbert2008} may still be affected by the limited resolution of the N-body simulation which, again, would  impact more smaller halos.
However, since we neglect the effect of baryons,
our result should be considered as conservative, as mentioned above. As noted in Figure 4 of \cite{Hilbert2008}, when baryons are added, the mass range that contributes the most to the optical depth, shifts from $\approx 8\times 10^{13}$ M$_{\odot}$ to $\approx 3\times 10^{13}$ M$_{\odot}$. \cite{Hilbert2008} concludes that baryons can increase the optical depth by up to a factor 2. As mentioned earlier, at low magnifications ($\mu<10$), the role of small halos is important so our estimation of the optical depth may be biased low by a factor $\approx 2$ in this regime. However, for the large magnifications we are interested in, the error introduced in the optical depth by ignoring the effect of baryons is expected to be less than that factor 2. 

Finally, as noted below in this paper, the baryonic component distorts the magnification through microlensing. As discussed in more detail below, microlensing reduces the maximum magnification of a small background source that could be attained  by a macromdoel. The reduction in magnifying power is proportional, to first order, to the surface mass density of microlenses, that is, to the baryonic component. Since around critical curves, the stellar component is in general larger for small halos than for massive halos, extreme magnification of small background objects is possible only when the stellar component near the critical curves is relatively small, that is in groups and clusters. 


\section{The role of microlenses}\label{sec_micro}

The above results are derived under the assumption that the caustic is not disrupted by substructure and that the $\mu^{-1} \propto \sqrt{|\beta-\beta_0|}$ law is maintained everywhere in the vicinity of the caustic in the source plane, where $\beta_0$ is the position of the caustic (in the image plane, the corresponding law is $\mu^{-1} \propto |\theta-\theta_0|$  where $\theta_0$ is the position of the critical curve). The probability density function (or pdf) of the magnification is simply $dN/d\mu \propto \mu^{-3}$. However, as discussed in \cite{Diego2018}, in the presence of substructure, including  microlenses with stellar or even substellar masses, the caustic can suffer severe distortion and the $\mu^{-1} \propto \sqrt{|\beta-\beta_o|}$  is no longer a valid approximation. The maximum magnification, that in theory can reach values as high as $\sim 10^7$ (for a background object with a size similar to the Sun) and a powerful lens such as a massive galaxy cluster, can be significantly smaller when microlenses are present. The maximum magnification is still proportional to the inverse of $\sqrt{R}$, where $R$ is the radius of the background object, but the proportionality constant is now related to the Einstein radius of the microlenses rather than the much larger Einstein radius of the macrolens. To be more precise, at low effective optical depth of microlensing\footnote{This definition was introduced in \cite{Diego2018} to account for the fact that a microlens in a region with magnification $\mu_{macro}$ and with mass M  behaves as microlenses with an {\it effective} mass M$\mu_{macro}$}, the proportionality constant is related to the Einstein radius of the microlenses, but magnified by $\sqrt{\mu_{macro}}$, where $\mu_{macro}$ is the magnification of the macrolens in the absence of microlenses \citep[see][for a discussion of this effect]{Diego2018}. When microlens are involved, we distinguish between the total magnification, $\mu$, (from the combined potential of the macromodel plus microlens) and the magnification of the macromodel, $\mu_{maro}$.

At high effective optical depth of microlenses even this scaling with $\sqrt{\mu_{macro}}$ breaks down due to overlapping microcaustics, and the maximum magnification becomes a function of the macromodel, surface mass density of microlenses and distance to the caustic of the macromodel \citep{Diego2018}. In this section, we explore the departure of the pdf from the standard law $dP/d\mu \propto \mu^{-3}$, when one approaches the caustic of a macromodel in the presence of microlenses. We make use of simulations similar to those described in  \citep{Diego2018} and the reader is directed to that reference for details on the simulations. Here, we simply summarize the most practical aspects of the simulation. For the macromodel, we chose a synthetic model that mimics the caustic region of a small galaxy. Adopting a relatively small galaxy is convenient, since the magnification drops more quickly as one moves away form the caustic that it would do if we were working with a caustic of a massive cluster. A rapid drop in the magnification allows us to study a wide range of magnifications in a relatively small area, and include regions in the source plane (relatively far away from the caustic) where the impact of microlenses is much less significant. The results derived in this section can easily be translated to larger lenses. For simplicity, we adopt a model that produces a vertical critical curve (and vertical caustic) so the magnification from the macromodel remains constant when moving in the vertical direction and falls as $1/|x-x_0|$ where $x_0$ is the position of the critical curve. The values of total $\kappa$ and $\gamma$ are $0.6667$ and $0.3333$ respectively at the position of the critical curve. For simplicity, we fix $\gamma$ and only modify $\kappa$ in either side of the critical curve with a slope of $10^{-7}$, which is appropriate for galaxy halos (the slope of $\kappa$ goes like the inverse of the Einstein radius of the lens in the vicinity of the critical curve for popular models like the isothermal model). 
The simulation covers a region of $0.178''\times0.00093''$ in the lens plane with a resolution of 31 nanoarcsec per pixel, resulting in approximately 1.7 billion pixels. The dimension in the horizontal direction is much larger than in the vertical direction. This is a consequence of the tangential magnification ($1/(1-\kappa-\gamma)$) being much larger than the radial magnification  ($1/(1-\kappa+\gamma)$). The much wider extension in the horizontal direction is required in order to capture the change in the properties of the magnification as one moves away from the caustic, but more importantly because the caustics from microlenses that are relatively far away from each other can still overlap in the source plane near the caustic. The deflection field from the microlenses is computed from a Spera model \citep{Spera2015} normalized to a surface mass density of $\Sigma_o = 19 {\rm M}_{\odot} {\rm pc}^{-2}$. This is similar to the value used in \cite{Diego2018} and a typical value in the outskirts of a galaxy. This surface mas density results in values of $\kappa _* \sim 0.008$ (or $\kappa _*/\kappa=0.012$), where  $\kappa _*$ is the convergence due to the microlenses. 

   \begin{figure}
   \centering
   \includegraphics[width=9cm]{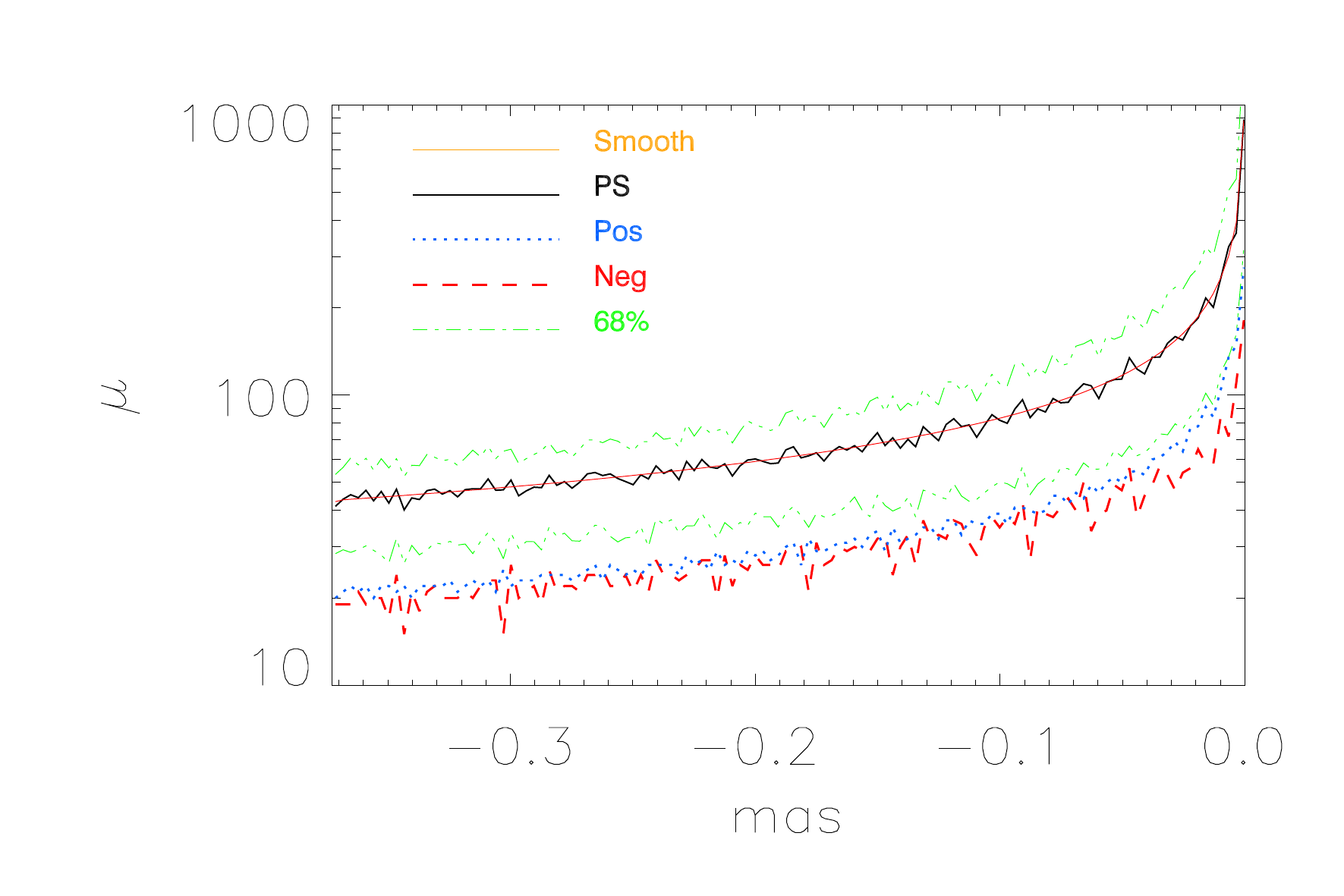}
      \caption{Mean magnification in the source plane and its variability as a function of distance to the macro-caustic.
              }
         \label{Fig_MicrolensVariance}
   \end{figure}

\begin{figure}
   \centering
   \includegraphics[width=9cm]{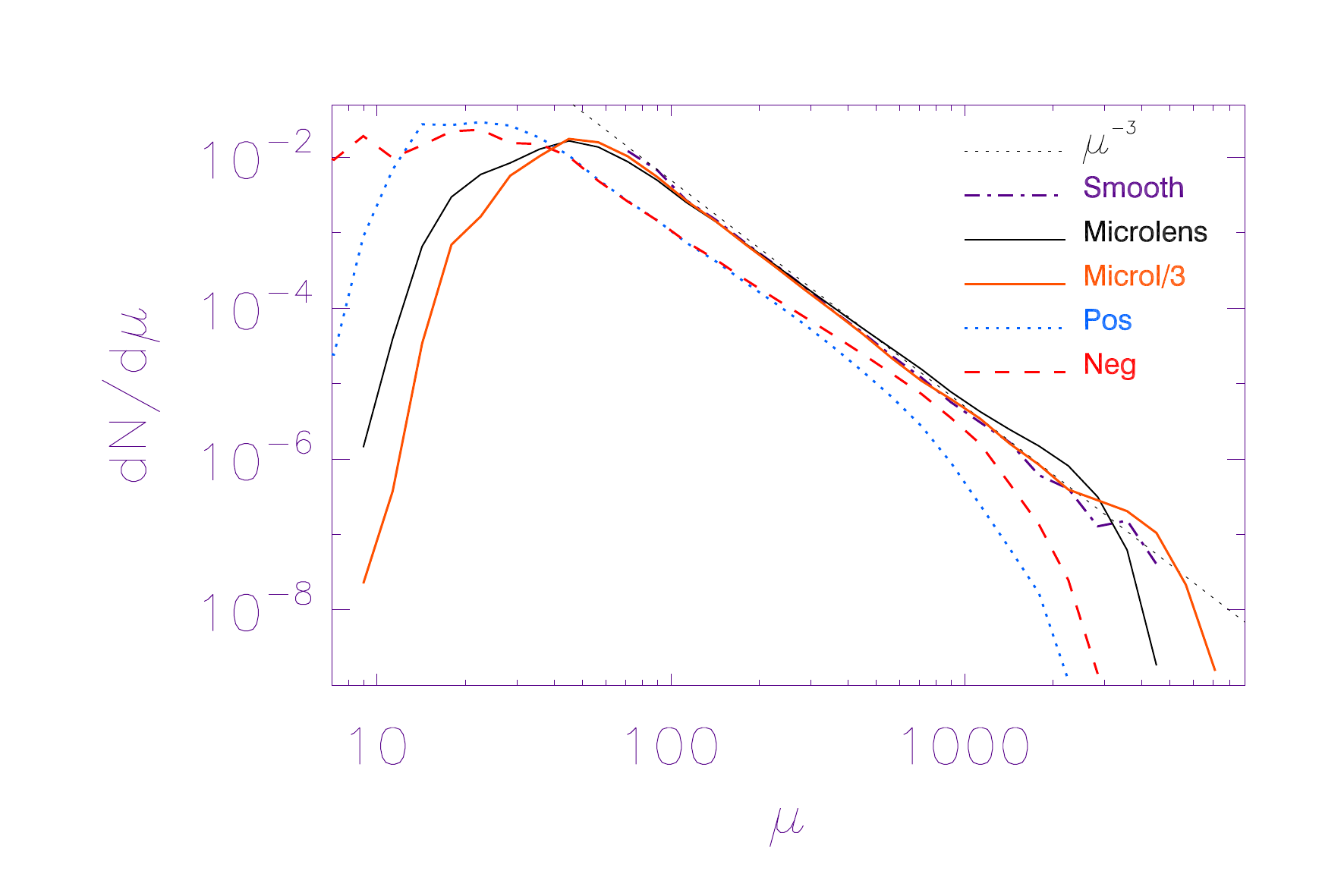}
      \caption{Probability of magnification in the source plane. The black solid line shows the probability of the total magnification when microlenses are present. The black dot-dashed line shows the probability computed in the same area but when microlenses are not included (i.e the smooth model). The black dotted line is the $\mu^{-3}$ law expected for smooth caustics. The blue-dotted and red-dashed curves are the probability of magnification from the regions with positive and negative parity respectively.
              }
         \label{Fig_MicrolensHisto1}
   \end{figure}

To compute the magnification in the source plane we use an inverse ray tracing algorithm. 
In Figure~\ref{Fig_MicrolensVariance} we show the average total magnification (black solid line) in the source plane after binning the source plane in regions of width $~3\,\mu as$ and as a function of the distance to the macromodel caustic. The red solid line shows the corresponding magnification of the macromodel, when microlenses are not included in the simulation. The average magnification is basically the same, reflecting the conservation of flux. The dot-dashed green line marks the 68\% region of the magnification variation. These curves lie very close to the dispersion values suggesting that, to first order, the pdf of the magnification when microlenses are present and at a fixed distance to the caustic can be approximated by a Gaussian. The width of this Gaussian scales with the effective surface mass density (that is, the surface mass density corrected by the factor $\mu_{macro}$). The dotted blue and dashed red lines show the average magnification as a function of the distance to the macro-caustic computed on the sides with positive and negative parity respectively. Each side carries roughly half the total magnification and behave very similarly in terms of average magnification at a fixed distance. However, this similar behavior is not maintained for the higher moments of the pdf. As discussed in earlier work \citep{Chang1979,Chang1984,Kayser1986,Schechter2002,Diego2018}, the pdf in the side with negative parity presents larger fluctuations than the corresponding pdf in the side with positive parity. The same result is found in our simulation, but covering a range in magnifications not studied in earlier work. Figure~\ref{Fig_MicrolensHisto1} shows the pdf of the total magnification (solid black) and also for the magnification in the sides with positive (dotted blue) and negative (dashed red) parities computed over a region of $370\times56$ $\mu$as$^2$ and including the macrocaustic.

The pdf of the total magnification when microlenses are included resembles the magnification of the smooth model except at the highest magnifications where, as discussed in \cite{Diego2018}, the maximum magnification is smaller than the corresponding maximum magnification of the smooth model. However, at more modest magnifications ($\mu \approx 1000$--$3000$), the model with microlenses has an increased probability compared with the smooth model. The contrary is found above $\mu > 2000$, where the probability of the magnification no longer follows $\mu^{-3}$ but falls sharply as a log-normal. Also, as discussed in \cite{Diego2018}, the side with negative parity exhibits a higher probability of large magnification than the side with positive parity. This is compensated by a larger probability of low magnification on the side with negative parity, confirming earlier results that show how fluctuations on the side with negative parity can be larger than on the side with positive parity \citep{Schechter2002}. The departure from the $\mu^{-3}$ law starts to be evident at $\mu \approx$~\,500 (below this value, a pdf evaluated over a wider region would show agreement between the smooth model and the model with microlenses at lower magnifications). The value of $\mu \approx$~\,500 above which the pdf deviates from the  $\mu^{-3}$ law can be approximately derived from theoretical arguments. Following  \cite{Diego2018}, the optical depth of microlensing, $\tau$, can be approximated by;

\begin{equation} 
\tau=\frac{\Sigma}{M}\pi\frac{4GM}{c^2}\frac{D_{ds}}{D_dD_s}\mu_t= 4.2\times10^{-4} \Sigma({\rm M}_{\odot}/{\rm pc}^2)\frac{a_2\mu_o}{\theta},
\label{Eq_tau}
\end{equation}
where $\Sigma({\rm M}_{\odot}/{\rm pc}^2) = 19$,  $a_2=1.5$ and $\mu_o=0.46''$ for the model considered in this work. $a_2=mu_r^{-1}$ represents the inverse of the magnification in the direction parallel to the macrocaustic, while $\mu_o$ defines the strength of the lens and enters in the expression $\mu=\mu_o/\theta$, where $\mu$ is the total magnification and $\theta$ is the distance to the critical curve of the macromodel. As discussed in  \cite{Diego2018}, the above expression for $\tau$ is derived under the assumption that all microlenses have the same mass. When compared with actual simulations based on a realistic mass function for the microlenses,  \cite{Diego2018} finds that Eq.~\ref{Eq_tau} overestimates the true optical depth by a factor $\approx 3$. Accounting for this factor 3 to correct the optical depth we find that $\tau$ saturates (i.e $\tau=1$) when $\theta=1.83$ mas. At this distance the total magnification from the macromodel is $\mu=\mu_o/\theta=250$ a factor $\approx$\,2 from the estimated value of $\mu\approx$~500, where the pdf of the total magnification for the model with microlenses start to deviate from the $\mu^{-3}$ law. 
The value of the magnification where this departure from $\mu^{-3}$ is observed is inversely proportional to $\Sigma({\rm M}_{\odot}/{\rm pc}^2)$. Figure~\ref{Fig_MicrolensHisto1} shows one more example corresponding to a situation where the  $\Sigma({\rm M}_{\odot}/{\rm pc}^2)=6.3$, that is, 3 times smaller than in the previous case. The pdf of the total magnification for this case is shown as a red solid line and the deviation from $\mu^{-3}$ takes place clearly at a higher magnification ($\mu \approx 2000$). Interestingly, the departure from  $\mu^{-3}$ is more abrupt, but the log-normal cutoff seems to be maintained. 
Smaller values of $\Sigma({\rm M}_{\odot}/{\rm pc}^2)=6.3$ are hence more favorable to observe more extreme events of many thousands but they have a smaller probability for intermediate magnifications of $\mu \approx 1000$. On the contrary, the larger the value of $\Sigma({\rm M}_{\odot})/{\rm pc}^2$, the sooner the pdf of the total magnification will deviate from $\mu^{-3}$ (although compensating the pdf of  $\mu$ with an increase towards magnification $\sim1000$). 

   \begin{figure}
   \centering
   \includegraphics[width=9cm]{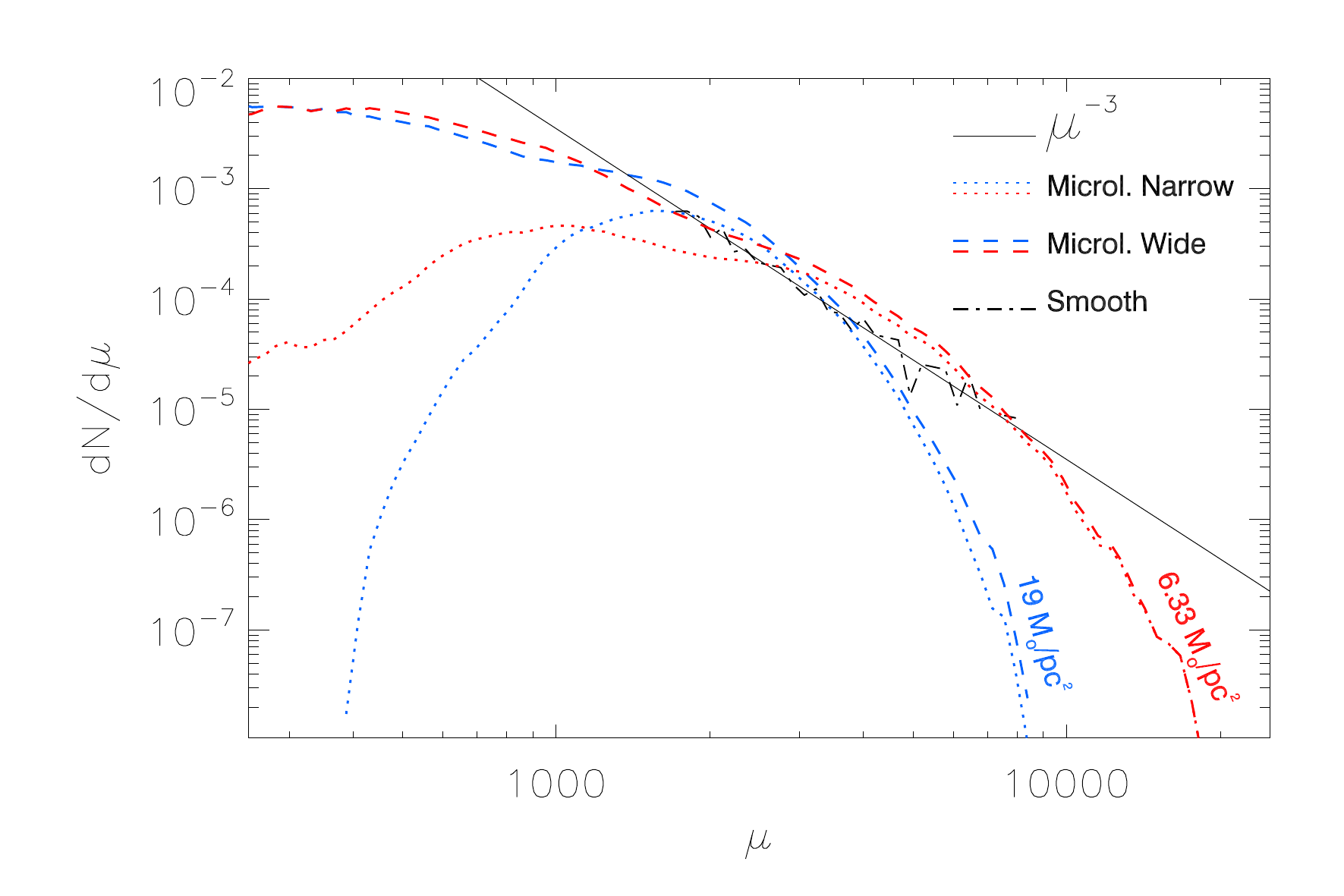}
      \caption{Pdf of the magnification computed in a smaller region around the macrocaustic at ten times the resolution. The black solid line shows the expected $\mu^{-3}$ for smooth caustics. 
               The black dot-dashed line shows the pdf of the smooth model computed through the inverse ray shooting method. The dotted lines are the pdf of the model with microlenses computed 
               over the narrow region in Fig.~\ref{Fig_MicrolensGlory} while the dashed lines are for the pdf computed over the wide region. Blue lines are for the model with 
               $\Sigma=19 {\rm M}_{\odot}/{\rm pc}^2$ and red lines are for a model with $\Sigma=6.33 {\rm M}_{\odot}/{\rm pc}^2$.
              }
         \label{Fig_MicrolensHisto2}
   \end{figure}

\begin{figure*}
   \centering
   \includegraphics[width=16cm]{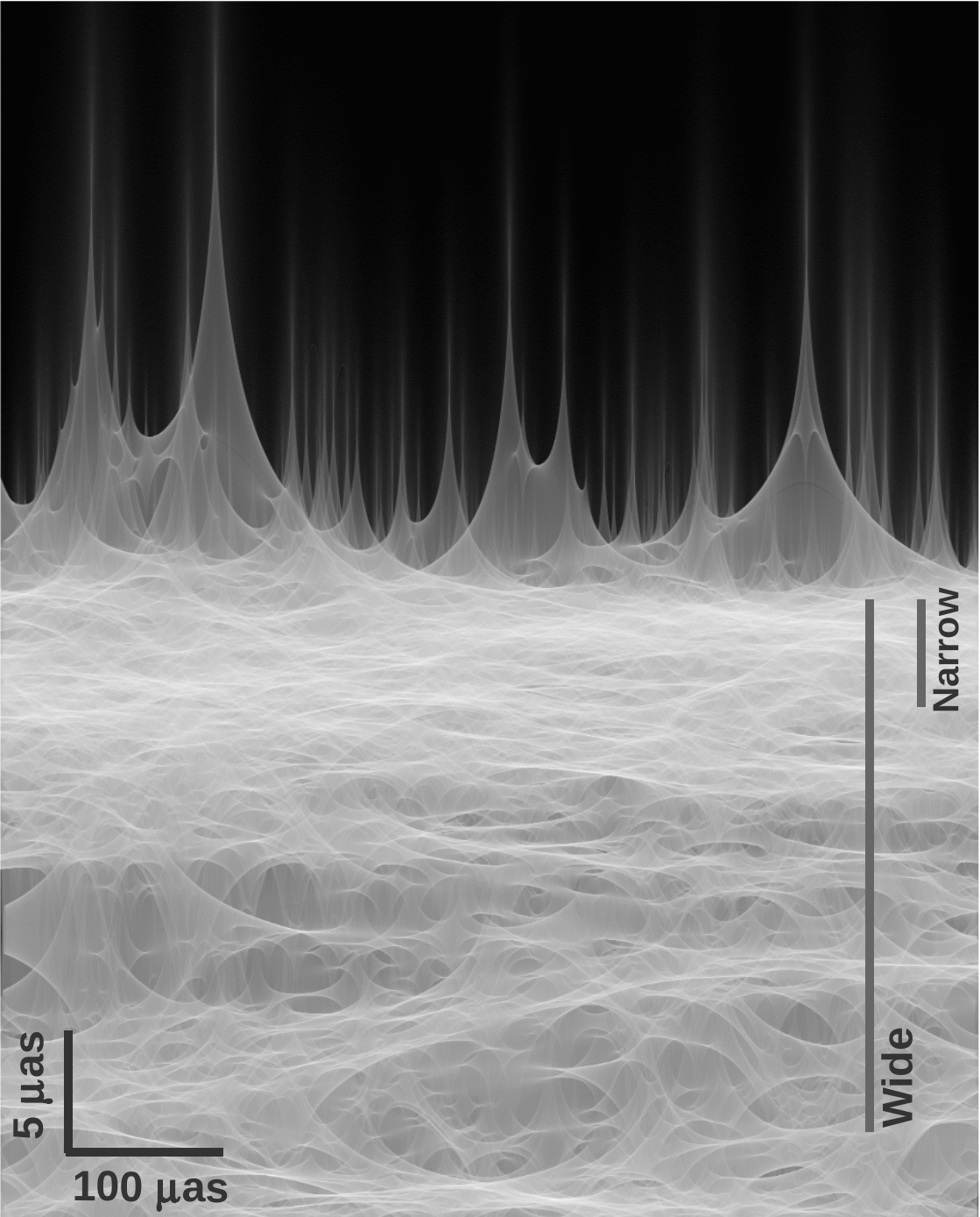}
      \caption{Magnification in the caustic region in the presence of microlenses with $\Sigma=19 {\rm M}_{\odot}/{\rm pc}^2$. The original simulated region has been compressed by a 
              factor $\sim 20$ in the horizontal direction in order to show a larger region. 
              The two vertical lines indicate the extension over which the pdf in Fig.~\ref{Fig_MicrolensHisto2} is computed for the narrow and wide regions. If microlenses were not present, the caustic drom the macromodel would be a single straight horizontal line with a region of zero magnification above it, and magnification decreasing as $1/sqrt{d}$ below it, where $d$ is the distance to the caustic.
              }
         \label{Fig_MicrolensGlory}
   \end{figure*}

In order to explore in more detail the departure from the $\mu^{-3}$ law we compute the magnification at a resolution ten times higher (that is, 3.1 nanoarcsec per pixel or $\approx 600$ times the diameter of the Sun at $z=1.5$) in a smaller region that zooms in around the macrocaustic. Since the maximum magnification scales as the inverse of the square root of the source radius, the results presented here are valid for sources with a radius $\approx\,1000$ times the radius of the Sun. A source with a radius, $R$, 100 times smaller than this could be amplified by a factor 10 times larger when its at a distance $R$ from the caustic. Although the simulated region is smaller, we still account for the contributions from the microlenses that lie far away from the macro critical curve. Figure~\ref{Fig_MicrolensHisto2} shows the pdf in this smaller region for the smooth model (black dot-dashed curve) and the model with microlenses (blue dotted and blue dashed curves). The difference between the blue-dotted and the blue-dashed is due to the different fraction of the caustic region that is used to compute the pdf. The blue-dotted curve is derived from a narrow region (width $\approx 5 \mu$as) very close to the macrocaustic while the blue-dashed curve is computed from a wider region (width $\approx 25 \mu$as). The widths of these regions, as well as the caustic zone, are shown in Figure~\ref{Fig_MicrolensGlory}, where for convenience we have rotated the simulated region, so the macrocaustic is oriented in the horizontal direction. 

The pdf in the narrower region (blue-dotted) follows the log-normal distribution originally discussed in \cite{Diego2018} that is typical of situations where many microcaustics overlap \citep[this is the saturation regime discussed in][]{Diego2018}. When computing the pdf over a wider region, the pdf starts to resemble the $\mu^{-3}$ law at lower magnifications ($\mu\sim2000$). The departure from this law at lower magnifications is largely due to the fact that the zoomed region does not include areas farther away from the macrocaustic that still contribute to these magnification bins. The black dot-dashed line shows the pdf of the smooth model (i.e, the same zoomed region but without microlenses). The black solid line shows the $\mu^{-3}$ law expected for smooth caustics. Note how the excess of probability (by a factor $\approx 2$) when microlenses are present is more evident in this plot around $\mu\approx 2000$. On the contrary, at $\mu\approx 10000$ the probability of magnification is at least two orders of magnitude smaller in the presence of microlenses, setting a natural limit to the maximum magnification around this value. If the contribution from microlenses is smaller, it is possible to increase the maximum magnification as shown by the red curves in Fig.~\ref{Fig_MicrolensHisto2}. These curves are derived from a model where the surface mass density of microlenses is three times smaller than the model in the blue curves. As shown in Fig.~\ref{Fig_MicrolensHisto1}, below $\mu\approx2000$, the pdf of the magnification follows closely the $\mu^{-3}$ law, but beyond  $\mu\approx2000$ the pdf deviates from this relation and at magnifications $\mu\approx10000$ the pdf is already falling like a log-normal. Since microlenses are expected to be ubiquitous up to the boundaries of most macrolenses, it is expected that observed extreme magnifications larger than a few tens of thousands will be extremely rare. 

If microcaustics are disturbed by microlenses and the maximum magnification is of order $10^4$ rather than $10^7$ as predicted for smooth caustics in clusters and small background sources like luminous stars, this may explain the apparent lack of observations of caustic crossing events from bright stars at $z=1$--$2$ \citep[with the only exception of the Icarus/Iapyx events discussed in ][]{Kelly2018}. To boost the flux of a distant luminous star by 10 magnitudes (so it can be detected) one needs a large magnifications of order $\sim10^4$ that, as discussed above, can be two orders of magnitude less likely when microlenses are present. Since smaller surface mass density of microlenses distorts the caustic less, the most extreme magnification factors are possible only with the smallest amount of microlenses. This implies increasing the redshift of the background source, so the critical curves move further away from the central region and the surface mass density of microlenses gets smaller. Alternatively, one can identify portions of the critical curves at lower redshift, which have relatively small contributions from microlenses. For example, the critical curves connecting two clusters in the process of merging produce near pristine critical curves in between the two clusters like the ones observed in the G165 cluster \citep{Frye2018,Griffiths2018}.

\section{Expected rates for compact bright sources}\label{sec_rates}
In this section we present rough estimates of the expected rates for different types of compact background sources. In particular we will focus on the expected rate of extreme magnification events (with $\mu >$\,100\,--\,1000) for, SNe, luminous stars at redshifts 1--2, Pop III stars at $z>6$, and gravitational waves from binary black holes (BBH), 
Rates of QSO will not be considered in this work has they have been studied in detail elsewhere (including the impact of microlenses). 
Also, since the accretion disc of a QSO is orders of magnitude larger, about 10 light-days (with 1 light-day $\approx 170$ AU), the maximum magnification for a QSO is of order $\sim 100$ times smaller \citep[see also][where it is shown how the size of the light emitting region can be significantly larger than this]{Blackburne2011,Guerras2013}.

SNe can reach also relatively large sizes (although not as large as QSO). More interestingly, the expanding photosphere from a SN explosion that is intersecting a caustic will exhibit very distinctive changes in the light curve as a consequence of the varying magnification \citep[see for instance the recent work of][and references therein]{Cobler2006,Goldstein2018,FoxleyMarrable2018}. 

During the early phases of the SNe (including the progenitor or pre-explosion phase), if the progenitor is at a fraction of 1 AU away from a caustic, the fainter but much more compact photosphere can be magnified by extreme factors owing to the relatively small size of the growing photosphere. After the initial explosion, the SN expands at rates of $\approx10^4$ km s$^{-1}$ for several days (or even weeks) resulting in sizes for the photosphere of $\approx 6$ AU one day after the explosion and $\approx 30$ AU after 5 days (and reaching $\sim 3\times10^{-4}$ pc at the time of maximum brightness). If the SN is located at a few tens of AU from a caustic (or microcaustic), its maximum magnification will peak when the photosphere touches the caustic. As the photosphere expands, the total magnification will start to decline leaving a distinctive signature in the light curve.

The probability of observing extremely magnified events at redshift $z$ depends on the product of the volume element at $z$, the volumetric density of objects (or events) at the same redshift and the probability of observing an event above a magnification $\mu$. Figure~\ref{Fig_MadauEventsGT100} shows the combined effect of the volume element, the evolution of the rates and the lensing probability. The dashed line shows the number of objects (or event rates) per redshift interval that trace the star formation rate as a function of redshift. For this particular example, we are using the intrinsic rate of SNe explosions (of all types). The solid line shows the number of these events  that would be magnified by a factor $\mu>100$. The dotted line shows the more realistic case, where observations are flux limited. We define the flux limit as the maximum distance at which an object would be observed without lensing, $D_o$. In particular, for this curve we adopt $D_0$ as the luminosity distance at $z=0.3$. Since the received flux scales as $D_l(z)^{-2}$, any event beyond $z=0.3$ brighter than the flux limit must have a magnification greater than $\mu = D_l(z)^2/D_o^2$ (the same law applies for gravitational  waves, where the signal-to-noise ratio of lensed objects scale as $\sqrt{\mu}/D_l(z)$). 

The location where the dotted and solid lines cross correspond to $\mu=100$, magnifications are larger than 100 to the right of this point and smaller than 100 to the left of this point. At low magnifications ($\mu<50$) the images may be resolved, and the total magnification may not be the best approach to describe the observations. Instead, one should consider the magnification carried by each image, which is $\approx 1/2$ the total magnification for $\mu$-factors larger than a few tens. Consequently, the dotted line at $z \approx 1$ (where the total magnification is $\approx 20$) would be overestimated by a factor $\approx 4$. 

\subsection{SNe at $z=$1--3}
\begin{figure}
   \centering
   \includegraphics[width=9cm]{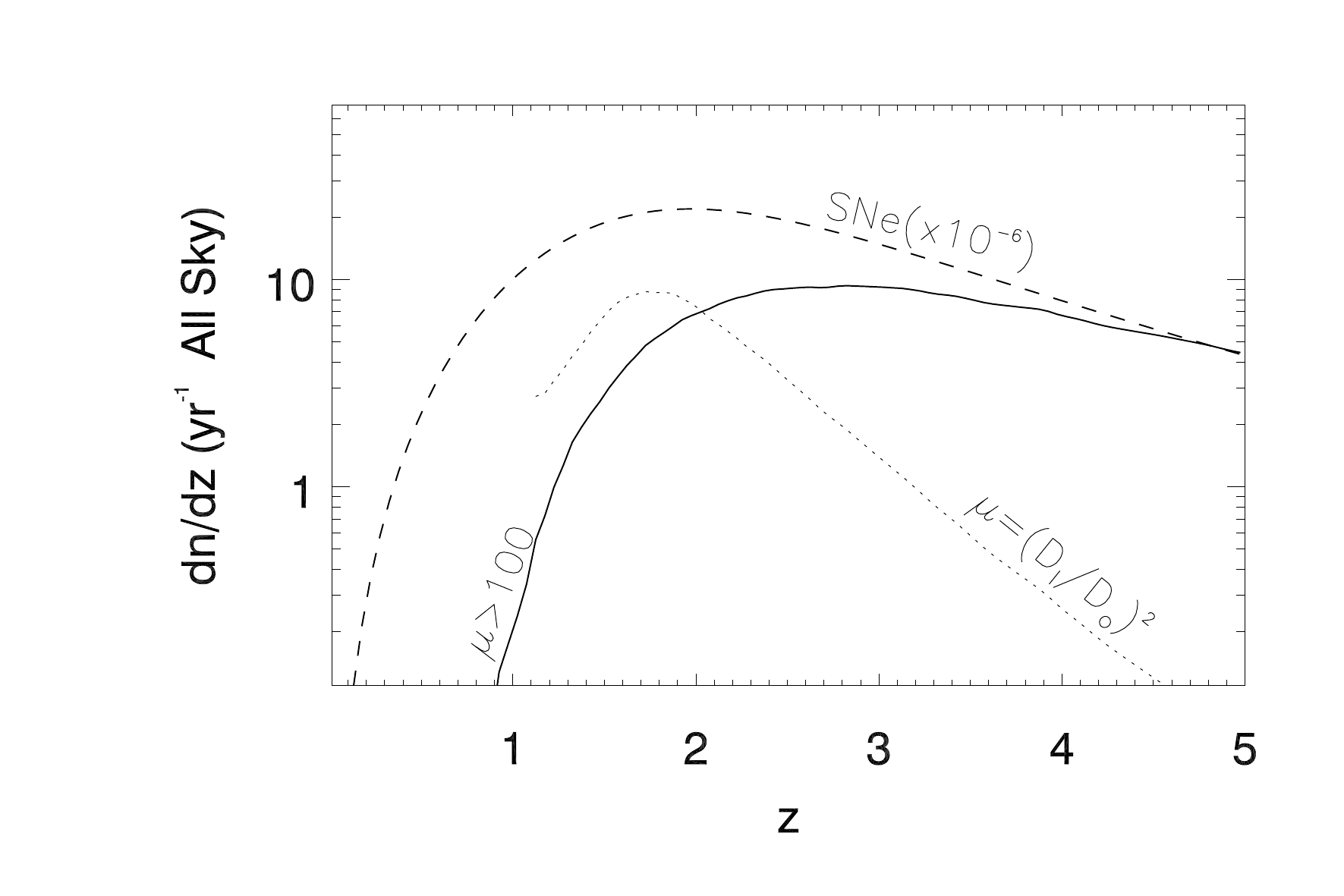}
      \caption{The dashed line represents the rate of events (all sky and in the source frame) for a model that traces the star formation rate and normalized to 
               $8\times10^{-5}$ events per year per Mpc$^3$. The SNe rate has been re-scaled by a factor $10^6$ for presentation purposes.
               The solid line, shows the product of the rate times the optical depth for $\mu>100$.
               The dotted line shows the rate of lensed events with the same apparent magnitude as a similar unlensed event at $z=0.3$.  
              }
         \label{Fig_MadauEventsGT100}
   \end{figure}

The volume in a shell of fixed thickness $\delta z$ is maximum at $z\approx 2.5$. The SFR density is maximum at $z\approx2$ \citep{Madau2014}. Hence, the volumetric rate of any event that traces the SFR will be maximum at around $z\approx 2$ as shown in Fig.~\ref{Fig_MadauEventsGT100}. The combined effect of maximal volume and maximal rate at $z\approx 2$ makes this an ideal redshift interval to search for events at extreme magnification. Despite the numerous detections of SNe, so far only three SNe have been gravitationally lensed with relatively large factors \citep[in the range of a few tens at most][]{Quimby2014,Kelly2015,Goobar2017}. In this subsection, we consider the case of SNe (of all types) at high redshifts ($z\approx$1--3) that are being lensed by extreme factors. As detailed below, during the much fainter phase of the first moments of the explosion, the SN is still small enough that magnifications of $\mu \sim 1000$ or larger can take place, if the SN occurs close enough to a caustic. 

After a SN explosion, there is typically an intense and short burst of flux lasting between a few seconds up to a few hours. This phase is known as the shock breakout. The breakout takes place $\approx 1.5$ hours after the explosion \citep{Garnavich2016,Bersten2018}. After $\approx 1$ day, the photosphere has grown by $\approx 1000$ solar radii assuming an expansion rate of $10^4$ km s$^{-1}$ \citep{Pejcha2015}. We focus on this period of 1 day  after the explosion (or 2--3 days in the observed frame for SNe at $z=$1--2), where the size of the SN is smaller than 1000  R$_{\odot}$ and the maximum magnification can reach factors $\mu>1000$. During this first day, the spectrum is concentrated mostly in the UV band. At $z=$1--2, a significant portion of this emission would be redshifted into the visible band making the SN visible in the more sensitive optical bands. For a SN like SN 2016gkg with absolute magnitude $-17$ at its maximum, the estimated magnitude before the shock breakout was $M_V\approx -12$ and reached $M_V\approx-15$ at the maximum of the shock breakout \citep{Bersten2018}. If such SNe can be observed without the help of gravitational lensing magnification at $z=1$ during its maximum (that is with apparent magnitude $m\approx 27$ attainable for instance in the deep-drilling fields planned for LSST), the same SNe, when observed amplified by a factor $\mu\sim 1000$ may reveal not only a much brighter main peak but also the shock breakout ($m\approx 25$) and even the last phase of the precursor ($m\approx 26.5$ ignoring k-corrections). 
 
To estimate the rate SNe at $z\approx$1\,--\,3 we adopt a model that traces the SFR \citep{Madau2014}. 
A model following the SFR is well motivated given the relatively short delay-time of $\approx 1$ Gyr (for type Ia) between the star formation and the SN explosion \citep{Ruiter2009}. 
We normalize the model to a rate of $0.8\times10^{-4}$ Mpc$^{-3}$ yr$^{-1}$ for all type SNe at z=0 \citep{Barbary2012}.  
For this model, the volumetric rate for any type of SNe at $z=1$ is $4.6\times10^{-4}$ Mpc$^{-3}$ yr$^{-1}$, and  $7\times10^{-4}$ Mpc$^{-3}$ yr$^{-1}$ at $z=2$ (both rates are given in the source frame, i.e rates are not corrected by time dilation). 
For type Ia SNe the corresponding rate would be about 3--4 times smaller \citep{Barbary2012}. 
Hence, the number of type Ia SNe explosion per year and redshift interval is estimated to be $\approx 10^7/2$ at $z\approx 1$ rising to $\approx 2\times10^7/3$ at $z \approx 2$ (see Figure~\ref{Fig_MadauEventsGT100}) where the dividing factors 2 and 3 account for the time dilation factor. 
For magnification larger than $\mu=1000$, the number of events per year is very small ($\approx 0.1$ event per year in the interval between $z=1$ and $z=3$). 
For more modest magnification factors, at $z=1$, we expect $\approx 0.5$ SN per year with magnification larger than $\mu=100$. At $z=2$ due to the increase in the intrinsic rate, volume element and optical depth, we expect $\approx 3$ SNe per year (where we have divided the intrinsic rate by a factor $1+z=3$). The increase in optical depth at higher redshifts partially compensates the decline in intrinsic rates beyond $z=2$, so the number density of lensed SNe with $\mu>100$ remains more or less constant between $z=2$ and $z=3$. 
Above $z=3$, the modest increase in optical depth is not enough to compensate the declining intrinsic rate and volume element per redshift interval. At magnifications $\mu=100$, the flux is boosted by 5 magnitudes, enough to see a sufficiently bright pre-shock phase with absolute magnitude $M_V=-12$ at $z=1$ with a telescopes like JWST. The shear number of events at $z\approx 2$ maximizes the probability that one of these events takes place very close to a caustic. 

Surveys covering a large region of the sky with good cadence should be able to identify these SNe as exceptionally luminous ones. One such survey is LSST which will monitor approximately 20000 deg$^2$ down to magnitude 24.5 ($5\sigma$ in r band) with a cadence of a few days \citep{Ivezic2008,LSST2009}. Hence, LSST should see $\approx2$ strongly lensed SNe ($\mu\sim100$ or more) per year between $z=1$ and $z=3$. 
The observed time delay between the multiple images can be used to independently determine the Hubble constant \citep{Vega2018}. 
Over the life time of LSST, there is a 50\% chance that at least one SNe has magnification larger than 1000, that is, the SNe is very close to a critical where the rapidly expanding photosphere will probe the structure of the overlapping microcaustics \citep[see for example][]{Schneider1986}. 

\subsection{SNe at $z=$6--12}
SNe can be extremely luminous and hence can be observed from large distances. Type Ia SNe have bolometric luminosities of order $10^{43}$ erg\,s$^{-1}$ \citep[see for instance][]{Sukhbold2018}, that is about $10^{10}$ more luminous than the Sun. Their corresponding absolute bolometric magnitudes is then $\approx -20$. The first type Ia SNe may have emerged after the formation of the first generation of white dwarfs. Since white dwarfs form from stars with a maximum mass of $\approx 10$ M$_{\odot}$, the minimum time to form a white dwarf is the lifetime of a star with $\approx 10$ M$_{\odot}$, that is only about 30 million years \citep[see for instance Table 1 of][]{Windhorst2018}. The delay time between the formation of the star and the SNe can be also relatively small and less than 1 Gyr as shown, for instance, by \cite{Maoz2010}. Hence, SNe Ia can appear very early on in the Universe, although it is no well known at what redshift the first SNe Ia may take place (after the formation of the white dwarf, the binary system needs to form). 

In this section we consider the case of type Ia SN at $z=6$--$12$ that can potentially be observed by JWST during the phase of maximum brightness and are relevant for cosmology (at least to test for possible changes in the equation of state of dark energy. Howevere, the reader should consider the rates assumed in this section as somewhat speculative, given the uncertainties in the formation rate of type Ia SNe at high-z and the unknown delay time between the formation of the star and the SNe which may significantly lower the rate at the highest redshift if the delay time is larger than 0.5 Gyr. Nevertheles, detection of such distant events through lensing could be used to infer the intrinsic rate (and delay times). Due to the larger size of the SNe during its maximum ($\sim 10^4 {\rm R}_{\odot}$), for a SNe sufficiently close to a caustic, the maximum magnification for a SNe can about an order of magnitude smaller than the magnification during the early phases of the explosion. 


For SNe at redshift $z>6$, JWST's NIRCam is, arguably, the best suited instrument to study these events. We consider the two wide filters F322W2 (covering the range $\sim$2.5-4 $\mu$m) and F444W  ($\sim$3.8-4.8 $\mu$m). F322W2 is more sensitive than  F444W, and it is ideal for SNe Ia at $z<10$, while F444W is better suited for SNe Ia exploding at $z>10$. At $z>12$, the peak emission in the redshifted spectra of the SNe moves to longer wavelengths than the ones covered by the F444W. At $z<8$, a SN explosion can be observed by JWST in the F322W2 filter even without the help of lensing. 

We assume an absolute magnitude of $M_V=-19$ during maximum for type Ia SNe \citep{Phillips1993,Brown2010}. The apparent magnitude is  given by $M_V + DM(z) + k(z)$ where $DM$ is the distance modulus and $k(z)$ is the k-correction. The distance modulus ranges from 48.85 at $z=6$ to 50.58 at $z=12$. 

For the k-correction we estimate $k(z)=-1.88$ for F322W2 at $z=8$ and $k(z)=-2.04$ for F444W at $z=12$\footnote{Patric Kelly, private communication}. 
Without any lensing, a SN at $z=6$ would have AB magnitude of 28 in F322W2 that could be detected with moderately deep observations. At $z=12$ a type Ia SN would have AB magnitude of 29.5, so in order to see it with reasonable integration times in the less sensitive F444W one would require lensing factors of 10 or more. 

To estimate  the rate of SNe at high redshift, we adopt a model that traces the SFR \citep{Madau2014}, but normalized to a rate of $1.25\times10^{-4}$ Mpc$^{-3}$ yr$^{-1}$ for type Ia at z=1 \citep{Barbary2012}. For this model, the volumetric rate of SNe Ia at $z=6$ is  $3\times10^{-5}$ Mpc$^{-3}$ yr$^{-1}$ and  $5\times10^{-6}$ Mpc$^{-3}$ yr$^{-1}$ at $z=12$. 
At the low end of the redshift interval, between z=6 and z=7 there are $365\times 10^9$ Mpc$^3$. The total number of SNe explosions in this redshift interval is $\approx 1.7$ million per year (after applying the time dilation factor), out of which only $\sim 1$ per year are expected to be lensed with total magnification $\mu>100$. 
At lower magnifications, $\approx 20$ SNe are expected to be lensed by a total magnification larger than 10, and with apparent magnitudes brighter than $m=25.5$ mag. Even though it is extremely unlikely that JWST will find one of these SNe, a wide survey like LSST can potentially detect some of these lensed type Ia SNe \citep[see][]{Rydberg2018}, that could be later followed up by JWST.


At the high end of the redshift interval, between $z=11$ and $z=12$ there are $\approx 225\times 10^9$ Mpc$^3$. The total number of SNe explosions in this redshift interval is $\approx 85000$ per year, out of which $\approx 1$ per 10 year are expected to be lensed by a factor larger than 100 (that is, apparent magnitude AB$\approx 24$ mag and hence detectable by LSST). The probabilities of detecting one of these SNe is small, but not negligible. A follow up with JWST of such SNe would provide valuable information about these first SNe.  

As mentioned in the beginning of this subsection, it is not well known what the rate of type Ia SNe may be at high redshift. Howevere, other SNe types, like core collapse, should be more abundant at $z>6$, and some of them could equally be observed through extreme magnifications.

\subsection{Luminous stars at $z\approx$1.5--2.5}
The small probability of observing high-z lensed SNe is partially a consequence of the relatively small number density of objects at high redshift. In order to compensate for the small probability of having a large magnification, one needs a large number of background objects so the probability of one of them being close enough to a caustic is larger. Bright stars (with absolute magnitude $M_V\approx -9$ or brighter) at $z\approx$1.5--2.5 (i.e with distance modulus in the range $\approx$45--47) could be observed with apparent magnitudes $\approx$27--29, if they are magnified by factors of a few thousand, hence within reach of future telescopes like JWST.
As shown in Figure~\ref{Fig_MadauEventsGT100}, the optimal range for detecting events or objects that trace the star formation rate is between $z=1.5$ and $z=2.5$. 

The mass-luminosity relation is well established for main sequence stars,
\begin{equation}
\frac{L}{L_{\odot}}=A\left( \frac{M}{M_{\odot}} \right ) ^{\alpha},
\end{equation}
where $A\approx 1$ and $\alpha=$3--4. For massive stars, the slope $\alpha$ is shallower \citep[$\approx 2.7$ for stars with $10< M_{\odot}<50$][]{Vitrichenko2007} and for the most massive and luminous stars the luminosity can reach the Eddington limit, that is $L \propto M$ \citep[see Eq. (3) and section 3 of][]{Windhorst2018}. Owing to the large value of the exponent $\alpha$, massive stars can be extremely luminous reaching luminosities several millions that of the Sun. However, at cosmic distances ($z>1$) and without the help of gravitational lensing magnification, their apparent magnitudes would be very faint ($AB>36$) and well out of the reach of current telescopes. With magnification factors of 1000 or more, some of these stars may get enough boost in their flux to be observable in deep observations. Small stars are more common and can be magnified by larger factors. The maximum magnification occurs when the star is at one radius distance from the caustic, that is $\mu_{max} \propto 1/sqrt{R}$ where $R$ is the star radius. Since the luminosity of a star, L,  scales as $L \propto R^a$, a lensed star has a maximum luminosity scaling as $L_{max} \propto R^{a-0.5}$. Since $a>1$, larger stars will always be more luminous at maximum magnification than smaller stars, and hence they are more likely to be observed during caustic crossing events despite being more rare.

The recent observation of the Icarus star at $z=1.49$ by \cite{Kelly2018} represents the first example of a redshift $z>1$ star magnified by factors of several thousand. \cite{Kelly2018} estimates an absolute magnitude for Icarus of $M_V=-9\pm0.75$ mag. At the redshift of Icarus (z=1.49) the distance modulus is 45.21 and the K-correction is estimated to be $-1.1$ magnitudes (in F125W) resulting in an apparent magnitude of $m=35.11\pm0.75$ mag (without magnification). The magnification for the Icarus event was estimated to be $\mu \approx$ 2000-4000 giving her a boost of 8.25--9 magnitudes, so that the event could be observed with $m=$26.11--26.86$(\pm0.75)$.\footnote{As this manuscript was finalzied, Icarus experienced a new microlensing event in early June 2018 with similar peak brightness as the original event in 2016.}

In this section we address the question of how common these events may be. In order to do this, we do a series of approximations adopting a conservative approach in those assumptions that are more uncertain.   
The number of bright stars at a given redshift can be estimated as a constant fraction of the number of stars in a galaxy times the number of galaxies.
\begin{equation}
N_{*}(L_{*},z) = \int dL N_{*}(L_{*},L)*\Psi_{gal}(L,z)*F_{SF}
\end{equation}
where $\Psi_{gal}(L,z)$ is the luminosity function of galaxies and $N_{*}(L_{*},L)$ is the fraction of stars with luminosity $L_{*}$ in the galaxy with luminosity $L$. 
The factor $F_{SF}$ accounts for the fact that massive stars have short lifes, so they will be predominantly found in galaxies where star formation is taking place. Elliptical galaxies for instance have small star formation rates, so the fraction of very luminous stars is expected to be much smaller.  
We set this fraction to $F_{SF}=0.5$ in the redshift interval considered. Observations show that the more numerous small galaxies ($M_*<10^{10} {\rm M}_{\odot}$) at $z=$1--2 are mostly star forming galaxies \citep{Ilbert2013,Muzzin2013,Ilbert2015}. 

For the luminosity function we adopt the best-fit Schechter parameters to the UV luminosity function in \cite{Reddy2009} at $z \approx 2$. After integrating the Schechter function between absolute magnitudes $-23$ and $-16$ we find a number density of $\approx 0.08$ galaxies per Mpc$^3$ (and a factor $\approx 2$ times less if the integral is carried out down to absolute magnitude $-17$). Assuming to first order no evolution in the luminosity function between $z=1.5$ and $z=2.5$ the total number of galaxies in this redshift interval is $\approx 4\times10^{10}$. 

To estimate the number of bright stars that may experience magnifications larger than 1000 (through temporary microlensing episodes), first we compute the number of galaxies that are intersecting a caustic. We assume that a galaxy is intersecting a caustic if its center is at a distance from the caustic such that the magnification of its nucleus is $\approx 100$. This means, the galaxy will produce multiple images (usually 3 or more), with two of the counterimages carrying most of magnification (i.e two images with magnification $\approx 50$ each at the nucleus position). This is a conservative assumption, since in most cases galaxies intersecting caustics are observed with average magnification $\approx$30--50 (like for instance the hist galaxy of Icarus). Since the average of the optical depth for $\mu>100$ between z=1.5 and z=2.5 is $\approx 10^{-7}$, we expect $\approx 4000$ galaxies intersecting caustics between $z=1.5$ and $z=2.5$. For each one, a portion of the galaxy will be at any time crossing a caustic. If a star is at a distance of a few parsecs from the caustic (i.e separations of 1 mas or less) where the magnification from the macromodel can reach factors of several hundred (see section~\ref{sec_math}) and microlensing may become important temporarily boosting the magnification to factors of a few thousand. At these magnifications, a very luminous star at $z\approx$1.5--2.5 could be elevated to magnitudes brighter than 28. If a galaxy of radius $R_{gal}$ is intersecting a caustic,  the probability that a given star in that galaxy is within a distance d (typically a few parsecs) from the caustic is $d/R_{gal}$, that is $\sim 10^{-3}$ for a galaxy of radius $\sim$few kpc. Hence, approximately one in a thousand stars in the galaxy has a chance of intersecting a microcaustic over the lifetime of JWST. If that galaxy has a number of super-luminous stars (with luminosities exceeding $10^6$ L$_{\odot}$) similar to our Galaxy (that is $\sim$100), this means every ten galaxies intersecting a caustic, one of them would have a super-luminous star close enough to the caustic to be observable.  This puts the number of background stars potentially observable at $\approx 400$. This is a very rough estimate and ignores many subtleties. 

An alternative, perhaps more precise, estimation can be obtained as follows. We assume that the number of bright stars in a given galaxy correlates with its luminosity. 
We assume that there are $n_*$ bright stars in a faint galaxy with absolute magnitude -16 (the number $n_*$ will be estimated later). 
The number of stars in a galaxy with absolute magnitude M will be then:
\begin{equation}
N_{*}(L_*,L) = n_*10^{-(M+16)/2.5}
\label{Eq_Nstar}
\end{equation}
where we have used the relation between magnitude and luminosity, $M - M_o = -2.5 log(L/L_o)$ with $M_o=-16$ and $L_o$ the corresponding luminosity of a galaxy hosting $n_*$ bright stars. The number of very luminous stars in our Galaxy and surroundings (LMC and SMC) with luminosities above $10^6$ L$_{\odot}$ or bolometric absolute magnitudes exceeding $-10$ mag is of the order of 100 
\citep[see examples in][]{Humphreys1979,Hamann2006,Crowther2010,Hainich2014,Crowther2016}. The total number is probably higher by a factor of a few but we adopt the conservative number of 100 for a galaxy like the Milky Way (plus its satellites).  
By requiring that Eq.~\ref{Eq_Nstar} reproduces the number of bright stars in our Galaxy (M=$-21$ mag, $z=0$, $N_{*} \approx 100$ bright stars), we find that $n_*\approx 2.5$. 
This is well motivated since the specific star formation rate of the Milky Way falls within the typical values for galaxies at z=0 \citep{PerezGonzalez2008,Licquia2015}.
We should note, however, that this simple recipe fails in cases like the large and small Magellanic clouds (LMC/SMC), where a significantly smaller number of bright stars would be expected while the number of observed extremely bright stars are as numerous in the LMC/SMC as in our own Galaxy. This may be consequence of interactions between the LMC/SMC and our Galaxy, which may trigger star formation episodes at the perigalactic passages of the LMC/SMC with our Galaxy \citep{Harris2004}. The expression above ignores such temporary episodes or interactions, and hence should be considered as a conservative approximation. 
After integrating over the galaxies brighter than magnitude $M=-16$ we find $\approx 1$ bright stars per Mpc$^3$, or $\approx 5\times10^{11}$ bright stars between $z=1.5$ and $z=2.5$ out of which $\approx 50000$ will be in areas where the magnification is larger than 100 and can momentarily undergo microlensing events by massive stars or BHs that boost the total magnification to $\mu>1000$, hence potentially observable by current telescopes. At macro-magnifications larger than 1000, the number drops by a factor 100, so we expect $\approx 500$ stars at macromodel magnification $\mu>1000$ between $z=1.5$ and $z=2.5$. In the context of the Icarus star with a macromodel magnification $\approx 500$ the number of bright stars (brighter than absolute bolometric magnitude $-10$ mag) moving in areas with macromodel magnification larger than 500 is 4 times more, that is $\approx 2000$ over the entire sky or $\approx 1$ stars per 20 square degrees. These stars, would appear very faint, but within reach of deep observations at AB$\sim$28--29 magnitudes. Occasionally, these stars may experience microlensing events where their flux may increase between 1 to 3 magnitudes. The duration of these microlensing events depend on the cluster microlens, the macromodel magnification, the relative velocity between the bright star and the caustic, and the angle between the star motion and the caustic orientation. For typical configurations, this timescale is of the order of days to weeks.  

The assumptions made above regarding the number of bright stars is quite uncertain\footnote{It should be noted that many of these extremely luminous stars can be also variable adding another element of uncertainty, since a star may be super-luminous only during a short period of time.}, but we should expect a reasonable number of stars that could be followed up. 
Icarus represents the first entry in a list of background stars at cosmic distances grazing in a field of microcaustics at high magnification. This list will grow quickly in the near future. Having a census of background stars that are moving near macrocaustics is of great interest, since it opens the door to study statistically the composition of the microlenses. Whether these microlenses are simply stars and remnants or more exotic microlenses like primordial black holes is something that could be determined with sufficient statistical information about the fluctuations in flux of the background stars (assuming time variability and microlensing features in the light curves can be distinguished). 

The planned Deep Drilling Fields of LSST \citep{Ivezic2008,LSST2009} will cover 38.4 deg$^2$ in four separate fields to a much greater depth (although how deep it is still undecided) and also better cadence. These fields are expected to reach magnitude 28 mag in r in the coadded data (and single-night depths of $\approx 26$ mag). At such depths, these fields should contain approximately two stars similar to Icarus. These stars will exhibit changes in their flux due to microlensing in time scales between days to weeks. They can also be later followed by JWST during one of the microlensing peaks to study their spectrum in detail.

Although not discussed in this work, observations of caustic crossings by Pop III stars will be possible also with future large telescopes like the ELT (in construction and expected to be operative in the middle of the next decade) or the proposed LUVOIR ($\sim$2035). 

   \begin{figure*}
   \centering
   \includegraphics[width=8cm]{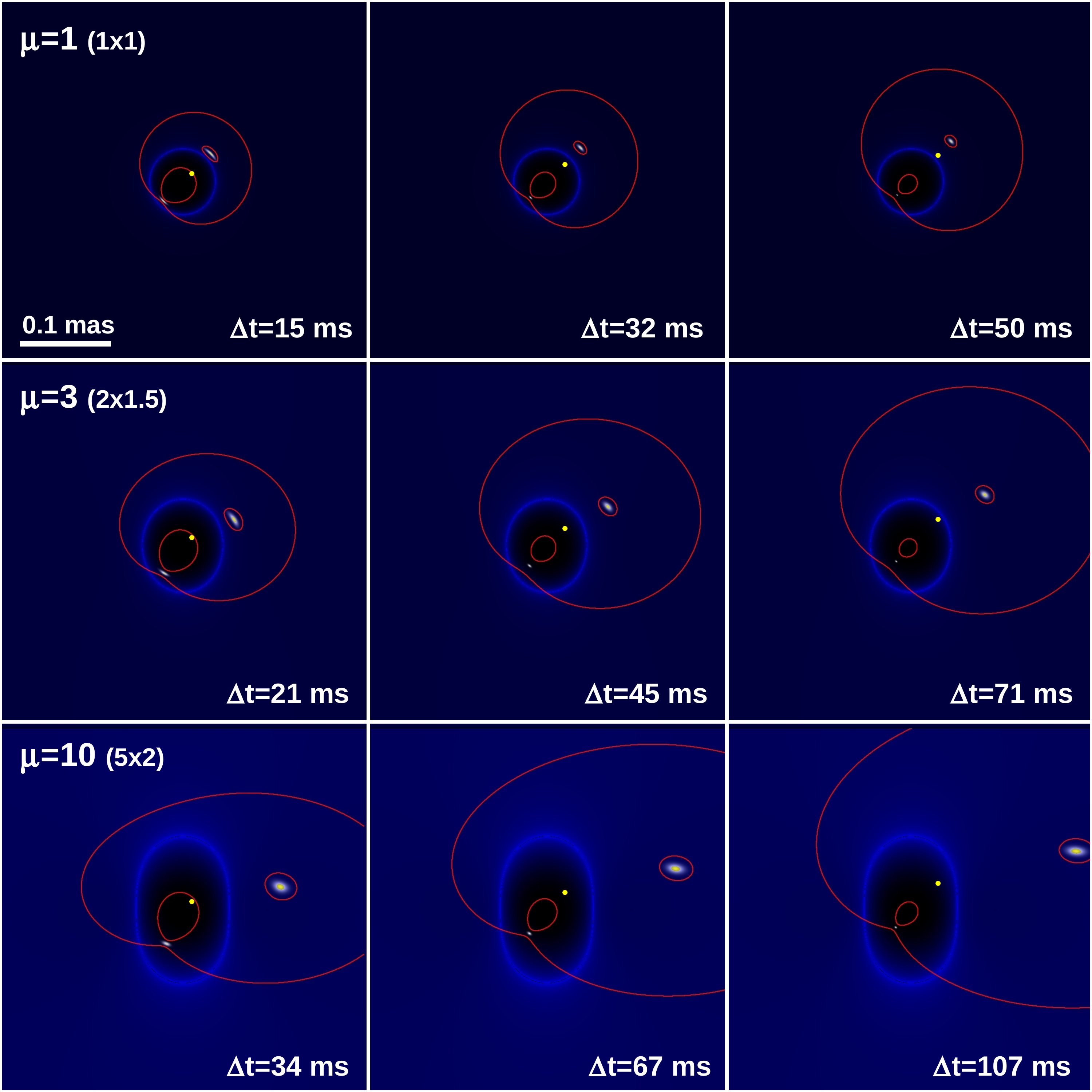}
   \includegraphics[width=8cm]{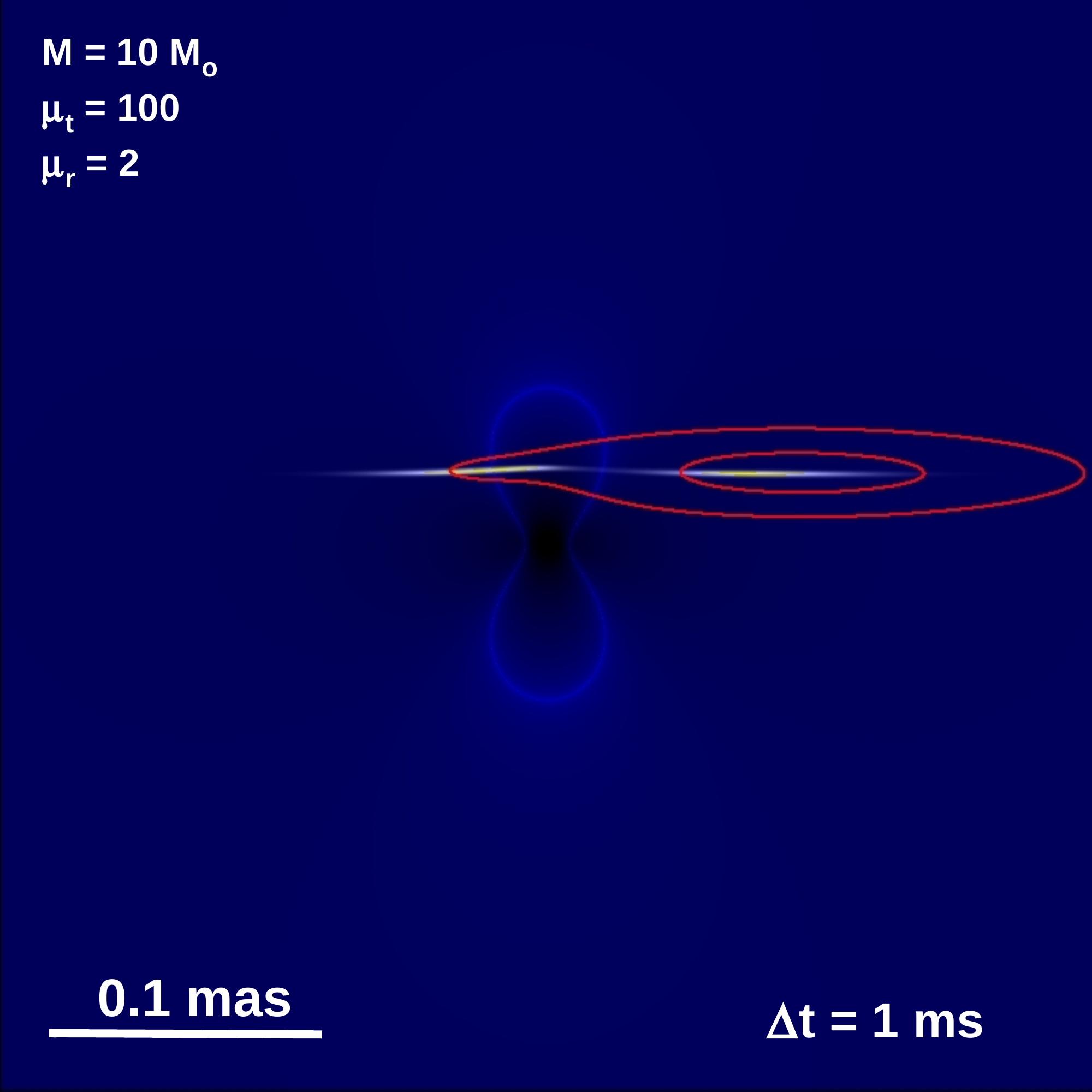}
      \caption{Time delay between the brightest counterimages (GW) produced by microlensing. 
               The group of 9 figures on the left is for a microlens of mass 400 $M_{\odot}$ 
               at $z=0.5$ and a source at $z=1.5$. Magnification in the lens plane is shown in blue/black. The critical curve can be seen as the bright blue contour. 
               The multiply lensed images are shown in white. 
               Each row corresponds to a different magnification.
               The numbers in the parenthesis next to the magnification indicate the tangential and radial magnifications respectively (i.e $\mu_t\times\mu_r$). 
               The microlens is isolated in the top row ($\mu=1$) and embedded in a potential from a macrolens with magnification 
               $\mu=3$ (middle row) and $\mu=10$  (bottom row). 
               Each column represents a different source position (marked with a small yellow dot). The red contours 
               are isochrones. The time difference between isochrones is given in the bottom right corner of each panel.     
               The right figure shows one example of a much lighter microlens (with mass 10 $M_{\odot}$) but much closer to the macrolens critical curve. 
               In particular, the macro-magnification at this position is 200 (100 in the tangential direction and 2 in the radial direction). 
               The redshifts are the same as on the left panels.  For this configuration, the time delay between microimages is of order 1 millisecond. 
               The source is much closer to center of the caustic, hence the smaller time delay. 
               Since the microcaustic is smaller, at separations from the center of the microcaustic similar to the ones in the left side, no multiple images 
               are produced.  
              }
         \label{Fig_TimeDelay}
   \end{figure*}

\subsection{Pop III stars}
At even higher redshifts, Pop III stars can be used as a powerful and very small background sources \citep{Bromm2001,Windhorst2018}. 
These stars are supposed to be extremely luminous owing to their large masses \citep{Bromm2001,Susa2014}, although recent work suggests that they may be less massive than previously thought \citep{Ishigaki2018}. 
The higher redshift has the advantage of having maximal optical depth, and the critical curves that form farther away from the centers of the lenses. At these separations, the impact of microlenses is smaller, thus allowing for larger magnification (at the expense of having fewer short episodes of microlensing). Despite having luminosities that can reach tens of millions $L_{\odot}$, the large distance makes their detection extremely challenging. Even deep observations made with powerful telescopes like the JWST may not reach the depth required to see them \citep{Rydberg2013}. However, thanks to gravitational lensing, some of these stars may get a boost of several magnitudes that could make them detectable. As discussed in \cite{Windhorst2018}, JWST's NIRCam will be able to look for highly magnified Pop III stars as caustic crossing events. JWST is expected to routinely reach AB$\approx$ 28.5--29 in medium-deep observations. In this section we present an order of magnitude estimation of the probability of observing these events. The number density of faint galaxies observed in deep observations (like the GOODS and HUDF field) is approximately between $\approx 1$ and a few galaxies per arcmin$^2$ in the redshift interval $z=$8--12 \citep{Finkelstein2015}. Similar results are derived when observing faint galaxies through powerful gravitational lenses like the Hubble Frontier Fields clusters \citep{Ishigaki2015}. The number of galaxies in the entire sky above $z=8$ (and within our horizon) can then be several hundred million galaxies and perhaps up to several billion. With such large numbers, even a small probability of intersecting a caustic results in a significantly large number of galaxies. 

From the previous section, microlenses negate the possibility of observing these stars at magnifications larger than $10^5$ (assuming that the observations are made in regions where the contribution form microlenses is small). For critical curves at high redshift ($z_s>7$), we expect the role of microlenses to be smaller than for sources at $z_s\approx$1--2. How large the maximum magnification is depends on how small the role of microlenses is at the position of critical curves for sources at high redshift. To estimate the surface mass density of microlenses at the position of a high redshift critical curve, we compute the critical surface mass density for a lens at $z=0.4$ and a source at $z=10$, $\Sigma_{crit}=1.8\times10^3 {\rm M}_{\odot}/{\rm pc}^2$ and assume that at the position of the critical curve, the convergence typically adopt values close to $\kappa\approx\,0.5$\footnote{Near critical curves, arcs typically have small radial magnification ($|1-\kappa+\gamma|\approx1$) and large tangential magnification  ($|1-\kappa-\gamma|\approx0$).}, that is $\Sigma = \kappa\times\Sigma_{crit}\approx 900 {\rm M}_{\odot}/{\rm pc}^2$. Adopting a mass-to-light ratio of 200 \citep[see for instance][]{Girardi2002}, we obtain that the surface mass density of stars is $\Sigma \approx 4.5 {\rm M}_{\odot}/{\rm pc}^2$, close to the low $\Sigma$ model adopted earlier with  $\Sigma \approx 6.33 {\rm M}_{\odot}/{\rm pc}^2$. Hence, if the surface mass density of microlenses at the position of high redshift critical curves is in the range of a few ${\rm M}_{\odot}/{\rm pc}^2$, we should expect maximum magnifications not exceeding a few tens of thousands. We adopt the value $\mu_{max}=5\times10^4$ as a reasonable limit (for a star with a radius several tens times the radius of the Sun, see typical radii for these stars in Tables 2-4 in \cite{Windhorst2018}. If we set the detection limit to $m=28.5$ mag, (a reasonable limit for medium-deep observations with NIRCam in JWST), this implies that only sources brighter than  $m=40.25$ mag can be observed through caustic crossing events. Following  \cite{Windhorst2018} section 4.4, we find the minimum mass that can be observed at this maximum magnification. These are are 70 ${\rm M}_{\odot}$, 190 ${\rm M}_{\odot}$ and 250 ${\rm M}_{\odot}$ at $z_s=7$, $z_s=12$ and $z_s=17$ respectively (these masses become 27, 36 and 42 respectively in the much shorter but also brighter phase of the asymptotic giant branch). Any mass below these limits would require a magnification larger than $\mu_{max}$.

\cite{Visbal2018} estimates that at $z<20$, the star formation rate density (SFRD) is already $\approx 10^{-4}$--$10^{-3} {\rm M}_{\odot}\,{\rm Mpc}^{-3} yr^{-1}$. A similar value is obtained by \cite{Sarmento2018}, but in the redshift range $z=$8--12. At these high redshifts, the SFR is expected to be dominated by the production of Pop III stars \citep[see for instance][]{Sarmento2018}. We make the simplifying assumption that most Pop III stars being produced at these redshifts are at a characteristic mass scale of $100\, {\rm M}_{\odot}$. This is close to the peak of the mass function found by \citep{Susa2014} for the first stars \citep[see also][]{Hirano2014}. 
This leads to  $\approx 10^{-6}$--$10^{-5}$ stars ${\rm Mpc}^{-3} yr^{-1}$ or 2--20 ${\rm Mpc}^{-3}$  during the typical lifetime of a massive Pop III star (t$_{mean} \approx 2$ Myr or $dz\approx0.05$ at z=10). This volumetric density translates into a surface density of $\approx$\,0.05\,--\,0.5 stars with $100\, {\rm M}_{\odot}$ per mag arcsec$^2$ between $z=10$ and $z=10.05$. 
This estimate is relatively uncertain, but is consistent with the observational limit in the near-IR of  $m\gtrsim31$ mag arcsec$^{-2}$ \citep[see][]{Windhorst2018}. Adopting the upper value (i.e 0.5 stars per arcsec$^2$ in $dz=0.05$ or 10 stars per arcsec$^2$ and redshift interval $dz=1$), the surface brightness from such density and redshift interval would be $m \approx 41$ mag arcsec$^{-2}$. After integrating over a redshift interval between z=7 and 20 we estimate the surface brightness would be  $AB \approx 36$ mag arcsec$^{-2}$, comfortably below  the limit $m \approx 31$ mag arcsec$^{-2}$. The number density of stars can be increased by a factor 100, and still not exceed this limit. Hence, as an upper limit we estimate a maximum number of stars of 1000 arcsec$^{-2}$ and per redshift interval at z=10. This estimate is in good agreement with the value estimated by \cite{Windhorst2018}, who estimate the surface density of bright (${\rm L}>10^{6}~{\rm L}_{\odot}$) Pop III stars at $z>7$ and brighter than AB $\approx 38.5$ mag to be in the range 1--1000 per arcsec$^2$ 

Adopting the compromise value of 10 stars per arcsec$^2$ or $\approx 5\times 10^{12}$ stars in the observable sky, the number of stars with $\mu>100$ would then be $\approx  5\times 10^{6}$ or $\approx 100$ per square degree. At $\mu>1000$, the number of lensed stars would be 1 per square degree. As a consequence, a surveyed area of 1 square degrees should contain at least one Pop III star at $z>7$ that is lensed by a factor $\mu>1000$ with $m<31$ mag, within reach of the JWST. The number density of extremely lensed Pop III stars can be increased by targeting known lenses at intermediate redshifts. \cite{Windhorst2018} estimates that the rate of extreme lensing events can be as high as 1 yr$^{-1}$ and per 3 massive lenses surveyed. As discussed in \cite{Windhorst2018}, Pop III stars are short lived but their remnants are expected to host bright accretion disc that could be equally lensed. These accretion discs are expected to live much longer than the Pop III stars, so their number density (and rate of lensed events) may be correspondingly higher.

\subsection{Gravitational waves}\label{sub_sect_GW}
\cite{Broadhurst2018} suggest that the low frequency events observed by LIGO can be reinterpreted as gravitational waves (GW) originating at $z>1$ that are magnified by factors $\mu>100$. \cite{Broadhurst2018} relies on the same lensing probability presented in this work, so we do not repeat the calculations presented in that work. Here, we briefly discuss their result and the impact of microlenses near critical curves in the context of GW, which was not explored in \cite{Broadhurst2018}.  
As proposed by \cite{Broadhurst2018}, the low frequency GW observed by LIGO would be produced by binary black holes (BBH) with intrinsic chirp masses of about 30 solar masses (which is a factor $\approx$\,2 times smaller than the values inferred by the LIGO collaboration) and that they would be redshifted so the observed chirp mass (derived from the observed GW frequency and its derivative) appears a factor two larger. In order for the strongly lensed GW interpretation to work in the context of the LIGO observations, the intrinsic rate of BBH coalescence needs to be $\approx3\times10^4$ events per year and per Gpc$^3$ and between $z=1.5$ and $z=2.5$. At this rate, the number of events in this redshift interval is $\approx 1.5\times10^7$ per year, out of which $\approx 4$ events per year will be magnified by factors above $\mu=100$ and hence are within reach of the LIGO detectors. Whether the low frequency events detected by LIGO are gravitationally lensed, as proposed by \cite{Broadhurst2018}, is something that can be tested with future data, since at these large magnifications one would expect to see multiple images of the same source with similar magnifications. \cite{Broadhurst2018} suggests that many of these multiple images are not observed since they fall within the detection limit of LIGO (due to the time delay between images and Earth's rotation). At the large magnifications considered in \cite{Broadhurst2018} ($\mu>100$), the effects of microlensing are expected to be important for small sources. However, given the long wavelength of gravitational waves one needs to consider wave optics and microlenses with masses below 30 ${\rm M}_{\odot}$ to produce negligible magnification of the GW. The maximum magnification of a GW by a lens of mass $M$ is $\mu_{max}\approx x/(1-e^{-x})$ where $x=(4/\pi)\times10^5 (M/M_{\odot})(\nu/Ghz)$ \citep{SchneiderBook}. For a typical GW detected by LIGO ($\nu\sim250$ Hz) and a microlens of mass $<10 M_{\odot}$, the maximum magnification is $\mu_{max}\approx 1$. At masses above 100 $M_{\odot}$ and frequencies of 250 Hz, the maximum magnification of GW scales as $\mu(M>100\,M_{\odot})_{max}\approx 3\times\,M/100\,M_{\odot}$. At higher frequencies or much larger masses, the exponential part tends to zero and $\mu_{max}\approx x$. Even though the magnification can be modest or negligible, there can be interference effects that can leave observable signatures even if the microlens mass is relatively small \citep[see for instance][]{Lai2018}. As discussed in \cite{Diego2018}, a microlens with mass $M$ near a caustic where the macromodel tangential magnification is $\mu_t$ behaves as a microlens with effective mass  $\mu_tM$. That is, at magnification factors of several hundred, a microlens with a mass of a few solar masses behaves as a microlens with an effective mass of several hundred solar masses. Hence, interference patterns (and some extra magnification) in the GW may occur if the GW is strongly magnified near a caustic, which is disturbed by microlenses.
Figure~\ref{Fig_TimeDelay} shows examples of time delays for a massive microlens of mass 400 $M_{\odot}$ (small panels on the left) and for a smaller microlens of mass $10 M_{\odot}$ (figure on the right). In both cases, time delays between 1\,--\,100 millisecond can be produced. Since the inverse of the frequency is also of order 1 millisecond, interference can take place in both cases. Moreover, if the lens plane is populated by many microlenses (as expected), \cite{Diego2018} shows how the number of multiple images can be larger, and also the time delay between them resulting in complex interference patterns. From the figures in the left panel, a trend is observed between the time delay and the magnification of the macromodel. For a fixed source position (with respect to the center of the microcaustic), the time delay grows as $\sqrt{\mu_{macro}}$. This means that relatively small microlenses sufficiently close to a critical curve can result in time delays which are large enough so interference can take place. On the other hand, as shown by \cite{Oguri2018}, the size of the microcaustics shows a weaker dependency with $\mu_{macro}$, so in order for these effects to be important with small microlenses, the separations between the source and the microcaustic must be small (almost independently of the macromodel magnification). Such smaller separations are required to produce situations like the one shown in the right side of Fig.\ref{Fig_TimeDelay}, where the probability of intersecting the smaller microlens is smaller than in the panels on the left side (with a larger mass and hence larger microcaustics). 
Hence, even though measurable interference patterns can be produced by stars of a few solar masses, their associated caustics are smaller. On the other hand, their larger number (compared with stars having $400 {\rm M}_{\odot}$) compensates for the smaller size of their caustics, thus increasing the probability of intersecting a small caustic. The superposition of multiple caustics helps also to increase the time delay, so microlenses near macromodel critical curves are expected to produce a rich phenomenology in terms of time delay and interference patterns. This regime is studied in more detail in \citep{Diego2019}.

Finally, GW from binary neutron stars (BNS) or neutron star and black hole (BNSBH) mergers can be equally magnified. In the case of BNS mergers, their larger number increases the probability of being lensed but the smaller masses limits the maximum distance at which BNS can be observed without lensing. Since the minimum magnification needed to detect a lensed GW effect scales as $(D_l(z)/D_o)^2$ (the strain o a lensed GW scales as $\sqrt{\mu}/D_l(z)$), to see an event similar to GW170817 (at a distance of 40 Mpc) but at a distance of 1000 Mpc ($z\approx 0.2$) one would require a magnification of $\mu \approx 600$. The optical depth at $z_s=0.2$ for  $\mu \approx 600$ is extremely small ($<10^{-8}$). At higher redshifts, the increase in volume and intrinsic rate of BNS mergers is not enough to compensate the rapid increase of magnification needed to promote the event above the detection threshold, so the prospects of seeing a lensed BNS are very dim (except for modest magnifications between $\mu=1$ and $\mu=2$, where there may be actually a chance of seeing such events). In the case of BNSBH mergers, we expect a similar scenario since, even though the mass of the BH component may be large, the intrinsic chirp mass remains small making the detectability of strongly lensed GW from BNSBH very unlikely. However, as noted by Diego et al. (2019), the smaller chirp mass produces strains with higher frequencies, and it is at these frequencies where the effects of interference from strlalr mass microlenses can be appreciated best.

Similarly, GW from SNe explosions are expected to be detectable only within our local neighborhood (tens of kpc) so the probability of them being lensed is even more remote.

\section{Discussion}\label{sect_discuss}
Massive stars are usually born after fragmentation of large clouds \citep{Tan2014}. This means that massive stars tend to form groups or small clusters \cite{Susa2014}. If one of these minihalos is close to a lens caustic, the probability of one of the stars in the group to cross a caustic grows compared with the case where luminous stars are isolated. A good example is the R136 group in the heart of the Tarantula Nebula (in the LMC) which contains several of the most luminous stars known. If one of these groups, with typical sizes of a few parsecs, is moving across a field of micro-caustics, each bright star in the group can undergo multiple microlensing episodes over a period of time greatly increasing the rate of events. This may create problems when interpreting the data since the number density of bright stars would be degenerate with the number density of microcaustics. In the starburst region of the Tarantula Nebula, \cite{Schneider2018} finds a flattening in the slope of the IMF at the high-mass end suggesting that very massive (and hence luminous) stars may be relatively more abundant than what would be inferred from standard IMF. Hence, the number of stars between $z=$\,1.5\,--\,3 with macromodel magnification larger than 500 could be more than our estimated number of 2000 based on standard IMF functions. 

We have considered also a different type of luminous source, SNe explosions. In particular, type Ia SNe are of great interest for cosmology since they can be used as standard candles to constrain the evolution of the equation of state of dark energy. Even though the explosions from SNe can release tremendous amounts of energy \citep{Woosley2002}, detecting them at redshifts larger than $z=$\,1--\,2 is challenging for current survey telescopes (typically with modest apertures) without the aid of lensing. In a recent work, \cite{Rydberg2018} consider strongly lensed distant SNe that could be observed by LSST. In their work, they also discuss the possibility of detecting Pop III SNe at $z<7$. In particular, they estimate that LSST may detect 1--2 Pop III between $z=5$ and $z=7$ in the deep survey (magnitude AB$\approx 28.5$ in 10 deg$^2$) although with a much more modest magnification factor than the ones considered in this work \citep[see also][for an estimate at lower redshifts]{Oguri2010a}. Studying in detail the light curves of strongly lensed SNe can be used to infer the underlying population of underlying microlenses \citep{Rauch1991}

At redshifts $z>7$, the first stars could be observed through caustic crossing events with telescopes like JWST as proposed by \cite{Windhorst2018}.  It is believed that at the end of their life, Pop III stars with masses in the range 15\,--\,40 ${\rm M}_{\odot}$ explode as core-collapse supernovae (CC SNe) and  with masses in the range 140–260 explode  as very energetic pair-instability  (PI)  SNe,  with luminosities up to 100 times higher than Type Ia or Type II SNe \citep{Heger2002} and leaving no remnant. At masses between 40\,--\,140 ${\rm M}_{\odot}$ and above  $\approx 260 {\rm M}_{\odot}$  the star collapses into a proto-neutron star or BH without producing a SN.  
Since SNe explosions from Pop III stars are considerably brighter than regular SNe, they can be detected by JWST \citep{Wise2005,Whalen2013a,Whalen2013b}. Both JWST and WFIRST should be able to detect SNe explosions from the first stars up to $z\approx 20$ \citep{Whalen2013b}. At lower redshifts (5\,--\,7), telescopes like LSST could also see lensed SNe from late Pop III \citep{Rydberg2018}.


An pointed out by \cite{Diego2018}, microlenses disrupt the caustics from macromodels and reduce the maximum magnification attainable by a small background source. The most favorable scenario to observe faint background sources is in regions around the lens where the contribution from microlenses is small. If the background source is at high redshift, the magnification required to compensate the increase in luminosity distance will be larger, but also the contribution from microlenses will be smaller, since the critical curves move outwards. Also, microlenses on the side of the critical curve with negative parity can magnify by larger factors at the expense of having longer periods of low magnification, where the background source apparently vanishes from the data. 
Due to microlenses, radial curves are less likely to produce extreme magnification, since the optical depth of microlensing is larger but on the other hand they are more likely to produce fluctuations, with more modest magnification factors of  $\mu\sim$~few tens or few hundreds. 
One can in principle estimate the maximum magnification expected at a given observed lensing arc based on an estimate of the surface mass density and a lens model. Similarly, if a given arc known to intersect a caustic is monitored for some time, one can infer the surface mass density of microlenses including non-luminous ones like remnants or primordial black holes (or PBH). The most favorable situation is for arcs intersecting caustics at high redshift. If one of these arcs is known to contain a sufficiently high number of bright sources, like for instance Pop III stars, and no extreme events are observed, one can directly use the lack of fluctuations to impose tight constraints on the abundance of PBH (assuming one is observing far enough from the lens's center so the contribution from the lens's stars and remnants is small, so a population of PBH would dominate the surface mass density of microlenses).  
Bright background sources at high redshift are more likely to undergo extreme magnification than the similar sources at low redshift, since the corresponding critical curves are observed in regions of the lens that are significantly less influenced by microlenses. Bright Pop III stars at high redshift are hence ideal targets for extreme magnification \citep{Windhorst2018}. 

Although not explored in detail in this paper, the role of microlenses can be important in the case of strongly lensed GW. If the volumetric rate of BBH mergers is of order $10^4$ per year and Gpc$^3$, we have shown how one should expect of order 1 GW per year magnified by large factors of 100 or more. At these magnifications, the amplification takes place near a critical curve where microlenses play an important role. We discussed how microlenses have a minor impact on the overall magnification of the GW, but for the right mass range, time delays of order 1 millisecond can produce interference patterns that could be observed both in the spectrum of the GW, but also as a different signal in the detectors, since they are probing different portions of the caustic region (although this second effect is expected to be much smaller than the former one). We have shown how in the case of a single microlens with a few solar masses near a critical curve, where the magnification can be more than 100, time delays of order 1 millisecond are possible. In a more realistic scenario, where the lens plane is populated by many microlenses, their corresponding microcaustics can overlap in the source plane resulting in larger time delays and complex interference patterns. 

\section{Conclusions}\label{sect_concl}
Giant arcs lurk around critical curves of massive galaxies, galaxy groups and clusters. These arcs can contain very luminous stars, or SNe and normally intersect the critical curves. If a super-luminous star, Pop III star, or SNe happens to lie at a few pc from the caustic, it can be magnified by factors of hundreds. At these magnifications, microlenses can distort significantly the pdf of the magnification (compared with the case without microlenses) producing temporary episodes of extreme magnification with $\mu>1000$. The length of these events depend on the relative velocity between the source and the caustic. The presence of microlenses also disrupts the caustic of clusters, preventing a caustic crossing event from reaching magnifications of millions. Instead, when microlenses are present, the maximum magnification of the caustic crossing event is in the range of tens of thousands. 
We compute the probability of extreme magnification with a state of the art mass function and analytical models for the individual lenses, which allow us to reach high spatial resolution. The magnification is computed with inverse ray tracing, the appropriate algorithm to account for the total flux of extremely-magnified, unresolved images. Our results are conservative since we ignore projection effects that increase the optical depth of lensing. We have also ignored the role of baryons that can increase the optical depth, specially from smaller halos, by up to a factor 2.
When microlenses are present, a trade off in the magnification between the macromodel caustic and the microcaustics results in an increase in the probability of magnifications around a few thousand, which can be a factor $\sim 2$, although the exact number depends on the macromodel and microlensing configurations. We find that the probability of having extreme magnification larger than 1000 for a source at $z=2$ is $\approx 3\times10^{-9}$. 
A source or event population with a volumetric density of $10^{-3}$ Mpc$^-3$ yr$^{-1}$ (comparable to the rate of SNe of all types) would then produce approximately one event per year with $\mu>1000$ at $z=2$. When accounting for microlensing near critical curves, this number can be increased by a factor of $\approx 2$. Among the most promising targets are Pop III stars for which we predict that as many as 1 star per square degree could have magnification larger than 1000. In combination with sporadic boosts from microlensing, these stars can be found with JWST in deep fields reaching $AB \approx 30$. Extremely magnified super-luminous stars at $z=$\,1.5\,--\,3 should be also relatively abundant. We predict that there should be $\approx 2000$ stars in the entire sky and in the redshift interval $z=$\,1.5--\,3 with macromodel magnifications similar to that of Icarus. The counterimages of these stars will be normally unresolved and affected by microlensing events. At typical relative velocities and microlensing surface mass densities, these events should take place every few years and with magnifications of a few thousand over periods of days to weeks \citep{Diego2018}, sufficient to be observed by future telescopes. 

Surveys like LSST will cover about half the sky with a cadence of a few days, sufficient to see strongly lensed SNe above z=1. We estimate about 5 strongly lensed SNe ($\mu>100$) per year above $z=1$ should be discovered by LSST. On the Deep Drilling Fields, we estimated LSST should detect approximately 2 stars similar to Icarus. These stars can be later followed up by JWST to get their spectrum and identify small fluctuations in the flux due to microlensing. 

If the rate of GW at $z>1$ is approximately one order of magnitude higher than the inferred rate from naively extrapolating the local rate, experiments like LIGO, that are instantaneously sensitive to the entire sky (neglecting the geometric factor), should routinely detect strongly lensed GW with magnification factors of 100 or more. At these magnifications, small microlenses with a few solar masses can trigger interference patterns in the GW that could be detected in their spectrum.

\begin{acknowledgements}
The author wish to thank Mitch Struble, Ravi Sheth, Gary Bernstein, and Ariel Goobar for very useful comments, Rogier Windhorst for carefully reviewing the manuscript, and Pat Kelly for important suggestions, as well as providing the K-corrections for JWST. The author wishes also to thank the referee for raising interesting questions about the manuscript. J.M.D. acknowledges the support of projects AYA2015-64508-P (MINECO/FEDER, UE), funded by the Ministerio de Economia y Competitividad. J.M.D. acknowledges the hospitality of the Physics Department at the University of Pennsylvania for hosting him during the preparation of this work.
\end{acknowledgements}

\bibliographystyle{aa} 
\bibliography{MyBiblio} 

\end{document}